\documentstyle[12pt,aaspp4,tighten]{article}

\input{psfig}

\def\etal{{et al.~}}

\def\VEV#1{\left\langle #1\right\rangle}
\def\abso#1{\mid\! #1\!\mid}
\def\la{\hbox{ \raise.35ex\rlap{$<$}\lower.6ex\hbox{$\sim$}\ }}
\def\ga{\hbox{ \raise.35ex\rlap{$>$}\lower.6ex\hbox{$\sim$}\ }}

\def\bfr{{\mbox{\boldmath $r$}}}
\def\bfk{{\mbox{\boldmath $k$}}}
\def\W2{{\cal W}}

\slugcomment{CAL-671, CU-TP-919, astro-ph/9903462}

\begin{document}

\title{The Power Spectrum, Bias Evolution, and the 
Spatial Three-Point Correlation Function}

\author{Ari Buchalter$^*$\altaffilmark{1} and Marc
     Kamionkowski$^\dagger$\altaffilmark{2}}
\affil{$^*$Department of Astronomy and Columbia Astrophysics
     Laboratory, Columbia University, 538 West 120th Street, New
     York, NY 10027}
\affil{$^\dagger$Department of Physics and Columbia
     Astrophysics Laboratory, Columbia University, 538 West 120th
     Street, New York, NY 10027}
\altaffiltext{1}{ari@astro.columbia.edu}
\altaffiltext{2}{kamion@phys.columbia.edu}


\begin{abstract}
We calculate perturbatively the normalized spatial skewness, $S_3$,
and full three-point correlation function (3PCF), $\zeta$, induced by
gravitational instability of Gaussian primordial fluctuations for
a biased tracer-mass distribution in flat and open
cold-dark-matter (CDM) models. We take into account 
the explicit dependence on cosmological parameters, the
shape and evolution of the CDM power spectrum, and allow the bias to be 
nonlinear and/or evolving in time, using an extension of Fry's (1996)
bias-evolution model. 
We derive a scale-dependent, leading-order correction
to the standard perturbative expression 
for $S_3$ (Fry \& Gazta\~{n}aga 1993) in the case of 
nonlinear biasing, as defined for the unsmoothed galaxy and dark-matter
fields, and find that this correction becomes
large when probing positive effective power-spectrum indices, i.e., scales
above $100$ $h^{-1}$ Mpc for reasonable CDM models.
This term implies that the inferred nonlinear-bias parameter, 
as usually defined in terms of the smoothed density fields,
might in general depend on the chosen smoothing
scale, and could allow better constraints on both 
the linear- and nonlinear-bias parameters 
on the basis of skewness measurements alone (or at least distinguish
between the smoothed and unsmoothed bias pictures), if 
$S_3$ could be measured over very large scales.
In general, 
we find that the dependence of $S_3$ on the biasing scheme
can substantially outweigh that on the adopted cosmology,
with linear and nonlinear bias separately giving rise to distinct signatures
in the skewness, but degenerate ones in combination. 
We demonstrate that the normalized 3PCF, $Q$, is an ill-behaved quantity,
and speculate that reported discrepancies between 
perturbative and $N$-body predictions for $Q$ 
may arise in part
from systematic errors associated with the poor choice of normalization.
To avoid this problem we investigate $Q_V$, the variance-normalized 3PCF.
The configuration dependence of $Q_V$ shows
similarly strong sensitivities to the bias scheme as $S_3$, but also
exhibits significant dependence on the form of the CDM power spectrum.  
Though the degeneracy of $S_3$ with respect to the
cosmological parameters and  
constant linear- and nonlinear-bias parameters
can be broken by the full configuration dependence of $Q_V$,
neither statistic can distinguish well between
evolving and non-evolving bias scenarios, since
an evolving bias is found to
effectively mimic a smaller but constant bias.
We show that this can be resolved, in principle, by
considering the redshift dependence of $\zeta$
which can also yield direct constraints on $\Omega_0$ and the
epoch of galaxy formation.

\end{abstract}

\keywords{large scale structure of the universe --- cosmology: theory --- 
galaxies: clustering --- galaxies: statistics}

\section{INTRODUCTION}

Recovery of the primordial distribution of density perturbations 
is crucial to understanding the origin of large-scale
structure.  Inflation, possibly the most promising paradigm for
the origin of structure, predicts Gaussian initial conditions (ICs),
while most alternative models generally predict some deviation
{}from Gaussianity.  It is well known that gravitational
instability will induce non-Gaussianity in an initially Gaussian
distribution of cosmological perturbations, as the nonlinear
collapse process gives
rise to interactions between initially uncoupled modes and thus to
non-zero higher-order moments.  As fluctuations
become nonlinear on progressively larger scales, larger
objects become virialized, producing a hierarchy of
structure.  In the quasi-linear (QL) regime, where the rms fluctuations
are small compared to unity, perturbative
solutions to the fluid equations for a self-gravitating
pressureless fluid in an expanding universe can be used to
derive the resulting $n$-point correlation 
functions\footnote{``Correlation function'' refers throughout to
the connected, or reduced, correlation function.} (CFs) and 
their Fourier-space counterparts, the
polyspectra, induced by gravitational instability
(Peebles 1980; Fry 1984; Goroff \etal 1986). In particular, 
the three-point correlation
function (3PCF), or equivalently, its Fourier transform (FT),
the bispectrum, is the lowest-order intrinsically nonlinear statistic and
can therefore place strong constraints on models of
structure formation\footnote{
Note that the $n$-point CFs are valid statistics only if the universe
is homogeneous on large scales (as opposed to exhibiting, say, 
multi-fractal structure). We invoke
the assumptions of large-scale homogeneity and isotropy, 
so that the CFs depend
only relative distances and are independent of orientation.}.

The perturbation-theory (PT) calculation predicts so-called
hierarchical behavior of the polyspectra, whereby
those of higher order can be simply related to the power spectrum,
since the physics on QL scales is still determined entirely
by the ICs. In the case of the bispectrum,
\begin{equation}
B({k}_{1},{k}_{2},{k}_{3}) = \mbox{${\cal Q}$}({k}_{1},{k}_{2},{k}_{3})
\left[P(k_{1})P(k_{2}) + P(k_{2})P(k_{3}) + P(k_{3})P(k_{1}) \right],
\label{B}
\end{equation}
where $P(k)$ is the power spectrum and
${ \cal Q}$ is of order unity and depends only on the magnitudes,
$k_i=\abso{\bfk_i}$, of the three wave vectors forming a triangle in $k$-space.
(Fry 1984; Goroff \etal 1986). 
A similar form was proposed,
on purely empirical grounds, by Peebles (1974, 1975, 1980)
for the spatial 3PCF,
\begin{equation}
\zeta(r_{12},r_{23},r_{31}) = Q(r_{12},r_{23},r_{31})
\left[\xi(r_{12})\xi(r_{23}) + \xi(r_{12})\xi(r_{31}) + \xi(r_{23})\xi(r_{31})
\right],
\label{zeta}
\end{equation}
where $\xi(r)$ is the two-point correlation function 
(2PCF) and the normalized amplitude, $Q$, 
depends only on the physical separations, $r_{ij} = \abso{\bfr_i - \bfr_j}$, 
of the three points in
real space given by coordinates $\bfr_i$, 
and is assumed to be of order unity. It should be noted
that though many authors assume $Q$ to be a constant, 
this is strictly untrue, nor is $Q$ given
by the $k$-space $\mbox{${\cal Q}$}$; rather it is a function defined
by equation~(\ref{zeta}), independent of the validity of the hierarchical
model.
A related, more tractable statistic which is commonly used is the normalized
spatial skewness, $S_3(R)$, related to 
$Q$ evaluated at zero lag for a distribution
smoothed over an effective scale half-width $R$.

Studies of the spatial 3PCFs of different tracer
populations on small scales (i.e., below 10 $h^{-1}$ Mpc)
have generally yielded results
consistent with the assumed hierarchical form above, i.e.,  
values of $Q \approx 1 \pm 0.5$ (Groth \& Peebles 1977;  Jing \& Zhang 1989;
T\'{o}th, Holl\'{o}si, \& Szalay 1989; Gott, Gao, \& Park 1991;
Jing, Mo, \& B\"{o}rner 1991; Baumgart \& Fry 1991; Bouchet \etal 1993;
Jing \& B\"{o}rner 1998)
and values of $S_3$ of order a few (Gazta\~{n}aga 1992;
Bouchet \etal 1993; Cappi \& Maurogordato 1995;
Gazta\~{n}aga, Croft, \& Dalton 1995), and numerical simulations have
also lent support to the hierarchical model (Bouchet, Schaeffer, \& Davis
1991; Fry, Melott, \& Shandarin 1993, 
1995; Bernardeau 1994a; Baugh, Gazta\~{n}aga, \& Efstathiou 1995;
Colombi, Bouchet, \& Hernquist 1996).
At larger scales, however, the detailed behavior of these higher-order 
statistics, 
such as the configuration-dependence of $Q$, is not well constrained.
Often, results for $Q$ are quoted as constant simply because
they reflect averages over all geometries, thus wiping out valuable
information. Jing \& B\"{o}rner (1997, hereafter JB97) and Matsubara \&
Suto (1994) do investigate the shape dependence of $Q$, but find substantial
disagreement between the predictions of QL PT and those of $N$-body
simulations. Jing \& B\"{o}rner (1998) further find significant devations
between the observed configuration dependence of $Q_{\rm proj}$, the
projected $Q$ perpendicular to the line of sight, and the
standard hierarchical result. These differences may be attributable
to various mechanisms, such as
nonlinear effects, finite volume effects, and projection effects, 
which can reduce the configuration dependence of both observed and 
$N$-body results
for $Q$, as compared with QL PT predictions. The observed discrepancies, 
however,
might also suggest that problems exist in the definition of $Q$.

That the normalized amplitudes $Q$ and $S_3$ have been observed
to exhibit roughly stable values near unity, as predicted, 
over scales on which $\zeta$ and $\xi$
individually vary by many orders of magnitude, has been interpreted
as loose support for
the scenario of gravitational evolution of Gaussian ICs. It is clear, however,
that the uncertainties in theoretical models, as well as in the
data, are still appreciable.
Much of this uncertainty is due to the fact
that existing redshift surveys have not been both large and deep
enough to yield strong constraints, particularly
on QL scales. However, 
with the advent of new, deep 
surveys at various wavelengths (radio, optical, infra-red, x-ray), 
probing many different populations
(galaxies, quasars, clusters, etc.), a rapidly growing database
of spectroscopic and photometric 
redshifts, and powerful new computing resources,
it will soon be possible to achieve precise measurements
of higher-order
spatial CFs. Interpreting the data
from these upcoming surveys will require more realistic modeling
of a variety of effects which may enter into the theoretical calculations.
For example, the 3PCF can
depend on the detailed form of the power spectrum 
(Frieman \& Gazta\~{n}aga 1994; JB97) and, particularly
for the case of angular CFs (see Buchalter, Kamionkowski, \& Jaffe 1999, 
hereafter BKJ), on the
density of nonrelativistic matter, the
cosmological constant, $\Lambda$, or the density of some other form of matter  
(Bouchet \etal 1992; Bernardeau 1994a; 
Bouchet \etal 1995; Catelan \etal 1995; Martel 1995; Scoccimarro
\etal 1998; Kamionkowski \& Buchalter 1998).
Furthermore, smoothing of the density field,
as occurs when determining $S_3$ from the moments of counts-in-cells, 
must be taken into account (Bernardeau 1994a,b).  
Perhaps most importantly,
however, biasing of the tracer population will have a
significant effect on the observed 3PCF (Fry 1984; Fry \& Gazta\~{n}aga 1993,
hereafter FG93).

Bias---the notion that observed structures do not exactly trace
the underlying density field---is advocated by both theory and
observations, and may in general be scale and/or time
dependent.  In deterministic models, 
the fractional density perturbation $\delta(\bfr)$ of
the unsmoothed tracer-mass distribution may be expanded in
terms of the perturbation $\delta_m(\bfr)$ to the total mass
distribution,
\begin{equation}
     \delta(\bfr) = b_1 \delta_{m}(\bfr) + \frac{b_2}{2} [\delta_{m}(\bfr)]^2
+ \cdots,
\label{dexp}
\end{equation}
where $b_1$ is the linear bias term, $b_2$ the first nonlinear
term, etc., and $\delta_m$ is itself written as $\delta_m = 
\delta_{m}^{(1)} + \delta_{m}^{(2)} + \cdots,$ 
where $\delta_{m}^{(n)} \ll \delta_{m}^{(n-1)}$, $\delta_{m}^{(1)}$ 
is the linear
solution, and $\delta_{m}^{(2)}$ is the leading-order 
departure from the Gaussian ICs. Note that the bias terms
defined above are scale-independent, since
equation~(\ref{dexp}) is a local relation between the 
unsmoothed density fields at every point (we 
comment further on this assumption below).
Like gravitational evolution or primordial
skewness, the existence of nonlinear bias will also give rise to
a non-zero third moment in the present-day density distribution
of the tracer mass (since the 3PCF is intrinsically second order,
and thus requires the $b_2$ term in the PT calculation). These effects
cannot be reliably distinguished by present data.
Furthermore, though Gaussian ICs are 
sufficient to produce hierarchical
behavior, they are not necessary; even if the data are shown
to follow the predicted hierarchy to high accuracy, it is possible
that this result is merely a coincidence of the bias. 
The impact of bias and its variation (thought to be
primarily with redshift) on the 3PCF must therefore be taken
into account if the predictions of Gaussian ICs
are to be tested unambiguously. It has been shown that
the detailed behavior of the 
3PCF can constrain the initial density profile (Fry \& Scherrer 1994;
Heavens 1998), 
distinguish between contributions to clustering from gravitational
collapse vs. bias (Szalay 1988; FG93;
Fry 1994; Frieman \& Gazta\~{n}aga 1994) and, for a given tracer population, 
yield determinations of the linear and nonlinear bias parameters
(Frieman \& Gazta\~{n}aga 1994;
Gazta\~{n}aga \& Frieman 1994; Jing 1997; Matarrese, Verde, \& Heavens 1997).
Moreover,
if different populations, such as clusters and galaxies, 
are differently biased relative to the underlying mass distribution, 
then measuring the 3PCFs of these
populations can provide multiple, independent constraints.
Several authors have recently argued that bias, and its evolution,
is the dominant consideration in evaluating structure-formation models
against clustering data, and have pointed to the need for employing more 
sophisticated bias models than the simplest 
ones often used (i.e., local, linear, 
and non-evolving) (JB97, Jing \& B\"orner 1998; Bernardeau 1995; 
Matarrese \etal 1997).

In this paper, we calculate the normalized spatial skewness,
$S_3(R)$, and normalized 3PCF, $Q(\bfr_1,\bfr_2,\bfr_3)$, in the
QL regime assuming Gaussian ICs. We consider open and flat cosmological
models with arbitrary allowed values of $\Omega_0$ and $\Omega_\Lambda$,
and take into account the dependence on the detailed
form and evolution of the CDM power spectrum.  
We further include the effect of
linear and nonlinear bias, and in particular, their time
evolution, via an extension of the Fry (1996) bias-evolution
model to the case of an arbitrary expansion history
(BKJ perform similar calculations to obtain
corresponding results for the angular skewness and 3PCF).  
For comparison, we derive the results for $S_3$ with a power-law
spectrum, including a scale-dependent, leading-order correction to the 
standard expression for $S_3$ (FG93)
in the case of a nonlinear bias as defined
in equation~(\ref{dexp}).
This correction term becomes appreciable for positive
effective spectral indices, corresponding to scales
$R \ga 100$ $h^{-1}$ Mpc for CDM models. The existence of this term
implies that the value of the nonlinear-bias parameter, as defined
for {\em smoothed} density fields, could in general
depend on the adopted smoothing scale. For a given CDM model,
this dependence, in principle, would allow a more accurate determination
of the linear- and nonlinear-bias parameters on the basis of skewness
measurements alone, provided however, that $S_3$ could be measured over 
sufficiently large scales.
We show that $Q$ is a poorly-defined statistic, exhibiting 
rapid variation and divergences which do not arise from the
behavior of the 3PCF itself, but are rather due to the quantity
$\left[\xi(r_{12})\xi(r_{23}) + \xi(r_{12})\xi(r_{31}) + \xi(r_{23})\xi(r_{31})
\right]$
in equation~(\ref{zeta}) acquiring values of or near zero in various 
cases [the $k$-space \mbox{${\cal Q}$} defined in equation~(\ref{B}) is
not subject to this normalization problem]. This dramatic 
behavior of $Q$ is in marked contrast to that predicted by some $N$-body
simulations (JB97; Matsubara \& Suto 1994) and
we propose that this discrepancy is due at least in part
to the practical inability to measure CFs
to high precision near zero. We
suggest that this may be addressed by considering $N$-body results
for $Q$ at earlier epochs, where nonlinear effects can be safely ignored;
to avoid this problem altogether, we define the variance-normalized
3PCF, $Q_V$,
given by $\zeta$ divided by the square of the variance at a given scale.
We find that $S_3$ and $Q_V$ do
indeed depend strongly on the bias scheme, as has been suggested.
In particular, a significant linear (nonlinear) bias
term produces a relative decrease (increase) in both normalized
amplitudes, and a
significant linear term is found to reduce the variation 
of $S_3$ with smoothing scale and of $Q_V$ with
triangle geometry, as compared with the unbiased predictions.
While $S_3$ is found to be only
mildly sensitive to the cosmological parameters, $Q_V$ can vary
appreciably with the value of the parameter
$\Gamma\simeq\Omega_0 h$, determined by the epoch of
matter-radiation equality, that locates the peak in the power
spectrum. In general, $Q_V$ is found to depend strongly on the
triangle geometry, with smaller, elongated structures,
corresponding to filaments or sheets, being more strongly
clustered.  
The rich configuration dependence of the 3PCF is shown to break
the degeneracy present in $S_3$ on scales $R < 100$ $h^{-1}$ Mpc, between
the cosmological parameters and in particular 
the constant linear- and nonlinear-bias parameters.  
Both statistics, however, are unable to
distinguish well between models of evolving and non-evolving
bias, since an evolving bias effectively acts as a smaller,
constant bias.  
We show that measurements of bias evolution, as well as of $\Omega_0$
and the epoch of galaxy formation, could potentially be
isolated by considering $\zeta$ as a function of redshift in a
very deep ($\bar z > 1$) survey.  Alternatively, BKJ show that
similar results can be obtained more feasibly from measurements of
the angular 3PCF.

\section{SPATIAL SKEWNESS}

The results below are based on leading-order PT and are thus
restricted to the QL regime, where $\xi \lesssim 1$.
Though the leading-order contributions
to higher moments of the density distribution are appreciable at
these scales, higher-order nonlinear contributions are not
expected to be significant, and various authors have
shown that observations and simulations in this regime are well-matched
by leading-order results alone (Szalay 1988; T\'{o}th, Holl\'{o}si,
\& Szalay 1989; Gott, Gao, \& Park 1991; Bouchet, Schaeffer, \& Davis
1991; Fry, Melott, \&
Shandarin 1993, 1995; Colombi, Bouchet, \& Hernquist 1996;
Scoccimarro \etal 1998).  In addition, numerical
simulations suggest that large scales still obey leading-order
PT even when small scales have become fully nonlinear (Bouchet
\& Hernquist 1992; Fry, Melott, \&
Shandarin 1995), and further that the results of
QL PT hold even on scales where the rms fluctuation is $\gtrsim
1$ (Bernardeau 1994a; Baugh, Gazta\~{n}aga, \& Efstathiou 1995;
Fry, Melott, \& Shandarin 1995).

We wish to derive an expression for the normalized 3PCF, $Q$, 
defined in equation~(\ref{zeta}), for a
distribution of tracer masses, such as galaxies, clusters, or radio
sources.  This quantity has the advantage that it is
independent of the overall normalization of the power spectrum.
The full spatial 3PCF is given by $\zeta(\bfr_1,\bfr_2,\bfr_3) \equiv
\VEV{{\delta}(\bfr_1){\delta}(\bfr_2){\delta} (\bfr_3)}$, where 
$\bfr$ is the comoving position
and $\delta(\bfr) = [\rho(\bfr,t_0) -
\overline{\rho}(t_0)]/\overline{\rho}(t_0)$ is the
present fractional perturbation to the unsmoothed tracer-mass density.
Although the spatial 3PCF is nominally the function of nine
quantities, statistical homogeneity and isotropy guarantee that
the 3PCF depends only on three parameters which may be taken to be, 
e.g., the three distances between the three points.  Still, evaluating
$Q$ over all allowed configurations 
may be fairly complicated, so it 
has become increasingly popular to focus instead on the 
normalized skewness $S_3 \equiv \VEV{\delta^3(r)}
\VEV{\delta^2(r)}^{-2}$. This 
statistic is likewise independent of the power-spectrum normalization
and is far easier
to calculate than its full $n$-point counterpart, while still preserving
much of the information contained therein.

In practice, $\delta(\bfr)$ cannot be evaluated continuously,
since the observed tracer-mass distribution is discrete, so we define
$\delta_R(\bfr)$ to be the density contrast smoothed over a cell with an
effective comoving scale half-width $R$.\footnote{
We implicitly assume $\delta = \frac{\rho}{\overline{\rho}}-1
\propto \frac{n}{\overline{n}}-1$, where $n$ is the number
of discrete counts.} 
In particular, we shall choose
a spherical top-hat window function; in terms of the Fourier
components, $\tilde{\delta}(\bfk)$, of the unsmoothed density field, 
\begin{equation}
    \delta_{R}(\bfr) = \int \frac{d^{3}\bfk}{(2\pi)^3}
     \tilde{\delta}(\bfk) e^{i \bfk\cdot\bfr} W(kR),
\label{dR}
\end{equation}
where $W(x) = 3 \sqrt{\pi/2} \: x^{-3/2} J_{3/2}(x)$ is the 
FT of the three-dimensional spherical top-hat window
function and $J_{n}(x)$ is a Bessel function of order $n$.
Since we are working in the regime where $\VEV{\delta^2} \ll 1$,
fluctuations at or below the smoothing scale
will not significantly affect our results (Bernardeau 1994a,b; 
see \S3).

The counts-in-cells spatial 2PCF is then given by
\begin{equation}
     \xi_R (r_{12}) = \VEV{{\delta_R}(\bfr_1){\delta_R}(\bfr_2)} = 
     \int\!\int
     \frac{d^3\bfk_1}{(2\pi)^3}\:
     \frac{d^3\bfk_2}{(2\pi)^3}\:
     e^{i(\bfk_1\cdot\bfr_1 + \bfk_2\cdot\bfr_2)} W(k_{1}R) W(k_{2}R)
     \VEV{\tilde{\delta}(\bfk_1)\tilde{\delta}
     (\bfk_2)}.
\label{xi}
\end{equation}
The power spectrum of the tracer mass, $P(k,t)$, is defined via
\begin{equation}
     \VEV{\tilde{\delta}(\bfk_1)\tilde{\delta}(\bfk_2)} \equiv (2\pi)^3
     \delta_D(\bfk_1+\bfk_2) P(k,t),
\label{P}
\end{equation}
where $\delta_D$ is the Dirac delta function and $k = \abso{\bfk_{1}}$.
Putting this into equation~(\ref{xi}), we have
\begin{equation}
     \VEV{{\delta_R}(\bfr_1){\delta_R}(\bfr_2)} = \int 
     \frac{d^3\bfk}{(2\pi)^3}\: e^{i(\bfk\cdot\bfr_{12})}
     P(k,t) W^2(kR),
\label{xi2}
\end{equation}
where one integral vanishes under the requirement $\bfk_1 = -\bfk_2$,
$\bfr_{12} = \bfr_1 - \bfr_2$, and we have made use of the the even
symmetry of the window function.
Taking $\bfr_1 = \bfr_2 \equiv \bfr$, we obtain an
expression for the variance,
\begin{equation}
     \xi_R(0)=\VEV{\delta_R^2(\bfr)} = \int 
     \frac{d^3\bfk}{(2\pi)^3}\: P(k,t) W^2(kR).
\label{xizero}
\end{equation} 
The linear solution to the equations of motion 
for the underlying density contrast has the separable form,
\begin{equation}
     \delta^{(1)}_m(\bfr,t) = D(t) \delta^{(1)}_m(\bfr,0), 
\label{dm1}
\end{equation}
where $D(t)$ is the linear-theory growth factor, given
(as a function of redshift) by
\begin{equation}
     D(z) = \frac {5 \Omega_0 E(z)}{2} \int_z^{\infty} dz^{\prime} \frac
     {1+z^{\prime}}{\left[E(z^{\prime})\right]^3}, \;\;\;\;\;\;\;\;
     E(z) = \sqrt{\sum_i \Omega_i (1+z)^{3(1+w_i)}},
\label{D}
\end{equation}
where $\sum_i \Omega_i = 1$ and $\Omega_i$ represents
the contribution to the overall energy density from species $i$ having
equation of state $p=w_i \rho$;\footnote{In this formulation, the curvature
of the universe contributes an amount $\Omega_K$ to
the total energy density, and yields the term $\Omega_K (1+z)^2$
in the sum in equation~(\ref{D});
Thus, in a universe with, for example, a ratio of nonrelativistic-matter 
density to closure density of $\Omega_0$ and 
a cosmological-constant contribution to
the total density of $\Omega_\Lambda$, $E(z)=\sqrt{\Omega_0(1+z)^3 
+ \Omega_\Lambda + (1-\Omega_0-\Omega_\Lambda)(1+z)^2}$.}
for an Einstein de-Sitter universe, $D(t)$ is simply the scale
factor, $a(t)$.
One can then see from equation~(\ref{P}) that the spatial
and time dependence of the power spectrum can likewise be separated.
Assuming the unsmoothed tracer-mass density contrast to be related to the
total-mass distribution via equation~(\ref{dexp}), we write the
leading-order result for the power spectrum of the unsmoothed
tracer-mass fluctuations as
\begin{equation}
     P(k,t) = A b_1^2 D^2(t) k^n T^2(k),
\label{Pkt}
\end{equation}
where $A$ is the overall amplitude and $T(k)$ is a model-dependent transfer 
function.
Substituting equation~(\ref{Pkt}) into (\ref{xizero}), setting
$x=kR$, and performing the 
angular integrations, we have
\begin{equation}
    \xi_R(0) = \VEV{{\delta_R^2}(\bfr)} = \frac{2 A b_1^2 D^2(t)}
     {(2\pi)^2 R^{n+3}} \int_0^{\infty} dx\: x^{n+2} T^2(x/R) W^2(x).
\label{xizerofinal}
\end{equation}

We now follow a similar procedure for the counts-in-cells 3PCF,
\begin{eqnarray}
     \zeta_R(r_{12},r_{23},r_{31}) & = &
     \VEV{{\delta_R}(\bfr_1){\delta_R}(\bfr_2){\delta_R}
     (\bfr_3)} = \int\!\int\!\int 
     \frac{d^3\bfk_1}{(2\pi)^3}\:
     \frac{d^3\bfk_2}{(2\pi)^3}\:\frac{d^3\bfk_3}{(2\pi)^3}
     e^{i(\bfk_1\cdot\bfr_1 + \bfk_2\cdot\bfr_2 + 
     \bfk_3\cdot\bfr_3)} \nonumber \\
     &   & \times \VEV{\tilde{\delta}(\bfk_1)\tilde{\delta}
     (\bfk_2)\tilde{\delta}(\bfk_3)} W(k_{1}R) W(k_{2}R)
     W(k_{3}R).
\label{zetaR}
\end{eqnarray}
The bispectrum, $B$, of the unsmoothed tracer-mass distribution is defined via
\begin{equation}
     \VEV{\tilde{\delta}(\bfk_1)\tilde{\delta}(\bfk_2)\tilde{\delta}
     (\bfk_3)} \equiv (2\pi)^3 \delta_D(\bfk_1+\bfk_2+\bfk_3)
     B({k}_{1},{k}_{2},{k}_{3},t),
\label{Bf}
\end{equation}
and is given, to leading order in PT, by (Fry 1984; Goroff \etal 1986;
Matarrese, Verde, \& Heavens 1997)
\begin{eqnarray}
    B({k}_{1},{k}_{2},{k}_{3},t) & = & 
    P(k_{1},t)P(k_{2},t)  \Biggl\{ \frac{1}{b_1}
    \left[ 1+\mu + \cos\theta\left(\frac{k_{1}}{k_{2}} + 
    \frac{k_{2}}{k_{1}}\right)
    + (1-\mu)\cos^{2}\theta \right]  \nonumber \\ 
    &   & + \frac{b_2}{b_1^2} \Biggr\} + \mbox{(cyc.)},
\label{Bf2}
\end{eqnarray}
where $\theta$ is the angle between $\bfk_{1}$ and $\bfk_{2}$, and $\mu$ 
is a function of the expansion history (Bouchet 
\etal 1992; Bernardeau 1994a; 
Bouchet \etal 1995; Catelan \etal 1995; Martel 1995; Scoccimarro
\etal 1998; Kamionkowski \& Buchalter 1998). For an Einstein-de
Sitter universe, $\mu=3/7$ (Peebles 1980).  Bouchet \etal (1992)
derive the approximation $\mu \simeq (3/7) \Omega_0^{-2/63}$
for an open universe.  Kamionkowski \& Buchalter (1998) find
$\mu \simeq (3/7) \Omega_0^{-1/140}$ for a flat universe
with a cosmological constant [see also Bouchet \etal (1995)].
Bernardeau (1994a), Catelan \etal (1995), and Scoccimarro \etal (1998)  
give 
analytical arguments that further show that $\mu$ will depend
only very weakly on the expansion history, and Kamionkowski \&
Buchalter (1998) verify this numerically for a variety of
models.  For plausible models with $\Omega_0 \geq 0.1$, the corrections to
$\mu$ are below $2$\% but we include them here for
completeness.  Note that the expression for
$B$ includes the dependence on the nonlinear bias term 
$b_2$, since the first
non-trivial terms in the connected $n$-point CFs require
PT to order $n-1$ [cf. equation~(\ref{dexp})].
Setting $\bfr_1=\bfr_2=\bfr_3\equiv\bfr$, $x_i = k_{i} R$, 
taking $\bfk_{1}$ to
lie in the $\theta = 0$ direction,
and inserting equations~(\ref{Bf}) and (\ref{Bf2}) into (\ref{zetaR}), 
we obtain the skewness,
\begin{eqnarray}
     \VEV{\delta_R^3(\bfr)} & = & \frac{6 A^2 b_1^4 D^4(t)}{(2\pi)^4 R^{2n+6}}
     \int_0^{\infty} dx_1 x_1^{n+2} T^2(x_1/R) W(x_1)
     \int_0^{\infty} dx_2 x_2^{n+2} T^2(x_2/R) W(x_2) \\
     &   &  \times\int_0^{\pi} d\theta \sin\theta  
     \left\{ \frac{1}{b_1} 
     \left[ (1+\mu) + \cos\theta\left(\frac{x_1}{x_2} +
     \frac{x_2}{x_1} \right) + 
     (1-\mu)\cos^{2}\theta \right] + \frac{b_2}{b_1^2} \right\}
     W(x_3), \nonumber
\label{skewness}
\end{eqnarray}
where a factor of 3 arises from symmetry considerations applied
to the 2 permutations in equation~(\ref{Bf2}). Noting that $x_3
= \sqrt{x_1^2 + x_2^2 + 2x_1 x_2 \cos\theta}$, we can evaluate
the $\theta$ integral by using summation theorems for Bessel
functions (Gradshteyn \& Ryzhik 1980; Bernardeau 1994a) which
lead to the following relations:
\begin{eqnarray}
     \int_0^{\pi} d\theta \: \sin^3 \theta \: W(x_3)  & = & \frac{4}{3} W(x_1) 
     W(x_2), \\
     \int_0^{\pi} d\theta \: \sin \theta \: (1 + \frac{x_2}{x_1}\cos\theta) \:
     W(x_3) & = & 
     \frac{2}{3} X(x_2) W(x_1), \\
     \int_0^{\pi} d\theta \: \sin \theta \: W(x_3) & = & \frac{4}{3} 
     \sum_{j=0}^{\infty} 
     (2j+3/2) W_{2j+3/2}(x_1)W_{2j+3/2}(x_2),
\label{3ints}
\end{eqnarray}
where $X(x)=3\sin{x}/x$ and
$W_n(x) \equiv 3 \sqrt{\pi/2}\: x^{-3/2} J_{n}(x)$, so that $W(x)=W_{3/2}(x)$.
Thus, we have
\begin{eqnarray}
     \VEV{\delta_R^3(\bfr)} & = &
     \frac{12 A^2 b_1^2 D^4(t)}{(2\pi)^4 R^{2n+6}} 
     \int_0^{\infty} dx_1 x_1^{n+2} T^2(x_1/R) W(x_1)
     \int_0^{\infty} dx_2 x_2^{n+2} T^2(x_2/R) W(x_2) \nonumber \\
     &   & \times \Biggl( \frac{1}{b_1} \biggl[\frac{1}{3}X(x_1)W(x_2) + 
     \frac{1}{3}X(x_2)W(x_1) \biggr] - \frac{2}{3}\frac{(1-\mu)}{b_1}
     W(x_1)W(x_2) 
     \nonumber \\
     &   &  + \frac{2}{3} \frac{b_2}{b_1^2} \sum_{j=0}^{\infty} (2j+3/2) 
     W_{2j+3/2}(x_1)W_{2j+3/2}(x_2) \Biggr).
\label{skewness2}
\end{eqnarray}
For a given choice of $b_1$ and $b_2$, the skewness obtained from 
equations~(\ref{xizerofinal}) and (\ref{skewness2}) effectively 
depends only on the smoothing radius, $R$ and the form of the
power spectrum; the explicit dependence on $\Omega_0$ and
$\Omega_\Lambda$ (through the dependence of $\mu$) is very weak.

\begin{figure}[htbp]
\plotone{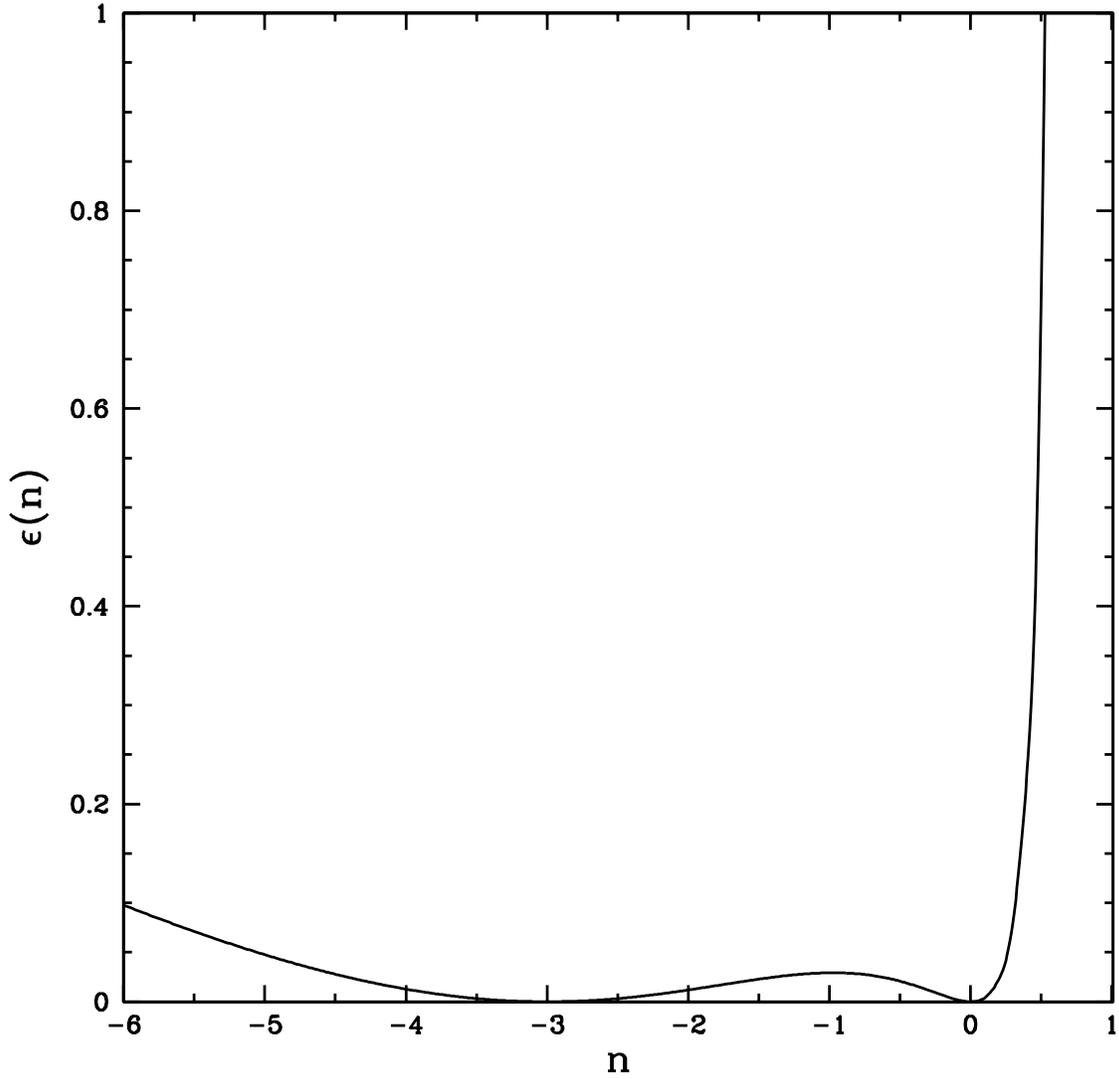}
\caption{The function $\varepsilon(n)$ from equation~(\ref{varepsilon}). 
For $-3 < n < 0$ the contribution of this term to $S_3$ in equation~(\ref{S3})
is negligible, but increases rapidly for $n > 0$, due to the increased
weighting of small-scale fluctuations by the unbounded power-law spectrum.}
\label{fig:varepsilon}
\end{figure}

\subsection{Effects of The Power Spectrum}

In general, we will consider a CDM transfer function given by
(Bardeen \etal 1986)
\begin{equation}
     T(q) = \frac{\ln (1+2.34q)/(2.34q)}{\left[ 1+3.89q + (16.1q)^2
     +(5.46q)^3 + (6.71q)^4 \right] ^{1/4}},
\label{Tk}
\end{equation}
where $q=k_p/ \Gamma$, $\Gamma \simeq \Omega_0 h$,
and $k_p$ is the physical wavenumber in units of $h$
Mpc$^{-1}$. The result for $S_3$ is then  obtained by using
equation~(\ref{Tk}) together with (\ref{skewness2}) and
(\ref{xizerofinal}). It is instructive to compare the CDM results
with those obtained assuming a power-law power spectrum,
$P(k) \propto k^n$. In this case, 
the $x$-integrals associated with the first two terms
in parentheses in equation~(\ref{skewness2})
can be evaluated analytically, and we obtain: 
\begin{equation}
     S_3 = \frac{\VEV{\delta_R^3(\bfr)}}{\VEV{\delta_R^2(\bfr)}^2}
     = \frac{1}{b_1}\left[\frac{34}{7}
     +\frac{6}{7}\left(\frac{7}{3}\mu-1\right) 
     - (3+n) \right] + 3\frac{b_2}{b_1^2}
     \left[ 1+\varepsilon(n) \right],
\label{S3}
\end{equation}
where
\begin{equation}
     \varepsilon(n) = \frac{2}{3} \sum_{j=1}^{\infty} (2j+3/2) 
     \frac{ \left[ \int_0^{\infty}
     dx \: x^{n-1} J_{3/2}(x) J_{2j+3/2}(x) \right]^2}{\left[ \int_0^{\infty} 
     dx \: x^{n-1} J_{3/2}(x) J_{3/2}(x) \right]^2}.
\label{varepsilon}
\end{equation}
Equations~(\ref{S3}) and (\ref{varepsilon}) are valid for $n$ in
the range $-3<n<1$. The individual terms in the
sum in equation~(\ref{varepsilon}) can be evaluated explicitly 
(Watson 1966), and we find that $\varepsilon(n) \ll 1$
for $-3<n<0$, but diverges for $n>0$, as demonstrated in Figure
\ref{fig:varepsilon}.
Like the $(3+n)$ term, the $\varepsilon(n)$ term [which
arises from the term proportional to $b_2$ in equation~(\ref{skewness2})]
reflects the effect of the smoothing filter on the biased, contiguous
field, $\delta(\bfr)$, defined in equation~(\ref{dexp}). The filter
is designed to separate smooth, large-scale fluctuations from the nonlinear,
small-scale field. However, a nonlinear distribution having a scale-free power
spectrum with $n>0$ contains so much small-scale power that the
deviations induced by smoothing over these small-scale fluctuations
become comparable to, or greater than, the perturbations above the
smoothing scale, thus leading to the divergence in $\varepsilon(n)$.

The form of equation~(\ref{S3}) agrees with 
the previously known scale-free result for $S_3$ for a constant, nonlinear 
bias (Bouchet \etal 1992; Juszkiewicz, Bouchet, \& Colombi 1993; FG93; 
Bernardeau 1994a; Fry 1994), with the exception
of the $\varepsilon(n)$ term. This difference arises from the fact
that other authors calculating higher-order moments of the smoothed
tracer-mass field (e.g., FG93) typically
define the bias parameters using 
the form of equation~(\ref{dexp}) to relate the {\em smoothed} tracer-mass
and total-mass fields, i.e.,
\begin{equation}
     \delta_R = b_{1}^{(s)} \delta_{m_R} + \frac{b_{2}^{(s)}}{2} 
     (\delta_{m_R})^2 + \cdots,
\label{dexpR}
\end{equation}
where the $b_i^{(s)}$ are the ``smoothed bias'' terms and
a subscripted $R$ again denotes a smoothed field.
Biasing and smoothing do not commute; whereas the unsmoothed bias parameters
we define in equation~(\ref{dexp}) are truly local and scale-independent,
the definition of bias in equation~(\ref{dexpR}) is inherently non-local,
only fixing the smoothed bias parameters
at the chosen smoothing scale. While it can be shown, using
perturbation theory, that $b_1^{(s)}=b_1$, 
the higher-order smoothed
bias terms will in general be different from their constant-valued
counterparts
in equation~(\ref{dexp}), and the magnitude of the difference will depend
on the scale on which the $b_i^{(s)}$ are defined.
Had we used the smoothed bias parameters as defined in equation~(\ref{dexpR}),
we would recover the result of FG93, i.e., equation~(\ref{S3}) with
$b_1 \rightarrow b_1^{(s)}$, $b_2 \rightarrow b_2^{(s)}$, and
$\varepsilon(n) = 0$. Comparing our result for $S_3$ with that of FG93, we
can infer that $b_2^{(s)}=b_2 [1+\varepsilon(n)]$. In the no-smoothing
limit, corresponding to $n\rightarrow -3$, 
we recover $b_2^{(s)}=b_2$, as expected.

Of course, the unsmoothed bias parameters are defined only in the fictitious
limit of continuous density fields. In practice, one might expect the
efficiency of galaxy formation to depend on the matter density not only
at the point of interest, but in the neighboring vicinity as well. Models
such as peak biasing and halo biasing (Mo \& White 1996) in fact imply
a biasing relation such as equation~(\ref{dexpR}), where the smoothed bias
parameters do in fact tend towards constant values in the large-scale limit.
Such a scenario would thus be inconsistent with the definition of the
unsmoothed bias parameters in equation~(\ref{dexp}), and would 
yield precisely the FG93 result for $S_3$, rather than equation~(\ref{S3}).
Fortunately, these two situations can be distinguished on the basis
of the large-scale behavior of $S_3$, as we will illustrate. 
Unless otherwise stated, it is hereafter understood that bias
shall refer to the relationship between the unsmoothed tracer-mass
and dark-matter fields, as given by equation~(\ref{dexp}).

\begin{figure}[htbp]
\plotone{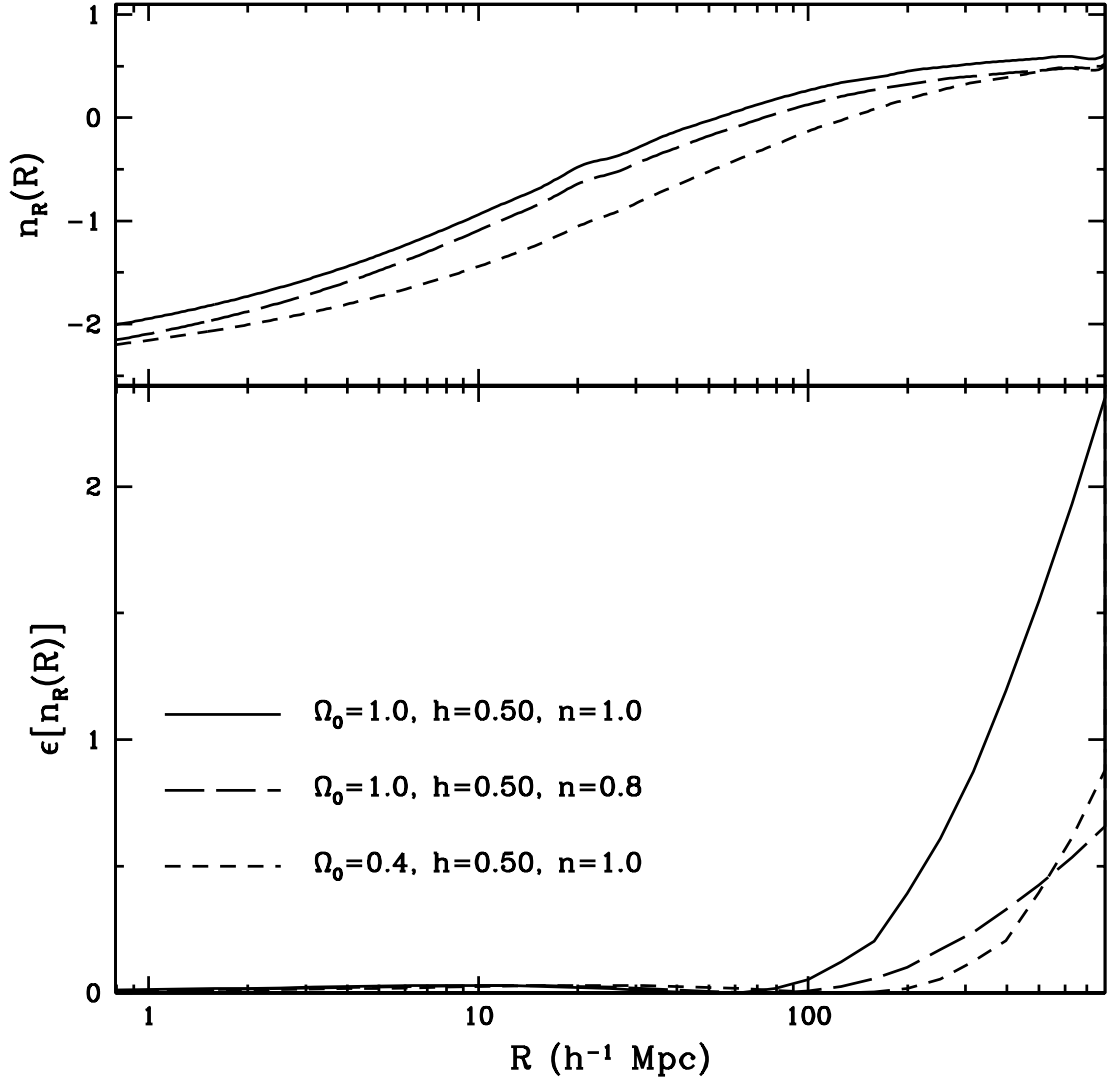}
\caption{The upper panel shows the results for $n_{R}(R)$ for
the three cosmological models described (the low-$\Omega_0$ result
is valid for both open models and flat models with a cosmological constant).
The lower panel shows the variation of the $\varepsilon$ term in
equation~(\ref{S3}) with smoothing scale in these CDM models. Note that this
term becomes comparable to unity on scales of a few hundred $h^{-1}$ Mpc,
where $n_{R}(R)>0$. The Figure should be interpreted only qualitatively,
since using $\varepsilon[n_{R}(R)]$ in equation~(\ref{S3}) appears
to overestimate the actual contribution of the higher-order
terms in the sum in equation~(\ref{skewness2}) for CDM models.}
\label{fig:epscorrection}
\end{figure}

Naturally, equation~(\ref{S3}) is $R$-independent, as expected for a
scale-free power spectrum. 
While the actual power spectrum for any tracer-mass distribution will not be 
simply a scale-free power law with $n>0$, it may be wondered
whether, in the unsmoothed biasing picture, 
the divergence seen in Figure \ref{fig:varepsilon} will affect
the results for $S_3$ on scales where the slope of the true (CDM) 
power spectrum is positive.
Bernardeau (1994a) shows
that the scale-dependent CDM result, without bias, can be recovered by using
an effective index,
\begin{equation}
n_{R} \equiv -\frac{d\log{\xi_{R}(0)}}{d\log{R}} - 3,
\label{ne}
\end{equation}
in place of $n$ in the scale-free result for $S_3$.\footnote{By 
contrast, using
an effective index defined by $n_{k} \equiv 
[d \log{P(k)}]/d\log k$, evaluated at wavenumber $k=c/R$
(where $c$ is constant of proportionality chosen as to yield the best-fitting
results), yields fractional errors in $S_3$ of at least 5\% (20\%)
for the smallest (largest) scales shown in Figure \ref{fig:epscorrection}.}
If we assume, for the sake of argument, that 
this holds true when calculating $S_3$ assuming our
nonlinear biasing model, we can illustrate the impact
of the $\varepsilon(n)$ term as a function of smoothing scale for various
CDM models.
The upper panel of Figure \ref{fig:epscorrection} shows $n_{R}(R)$ for
for three flat ($\Omega_0+\Omega_\Lambda=1$) 
cosmological models: standard cold dark matter (SCDM; $\Omega_0=1$, 
$h=0.5$, $n=1$),  
a tilted CDM model (TCDM; $\Omega_0=1$,
$h=0.5$, $n=0.8$), and CDM with a cosmological constant 
($\Lambda$CDM; $\Omega_0=0.4$, $h=0.5$, $n=1$), 
while the lower panel shows $\varepsilon[n_{R}(R)]$ for these models.
One can see from equations~(\ref{xizerofinal}), 
(\ref{varepsilon}), and (\ref{ne}) that the results from the $\Lambda$CDM
model will be identical to those for an open
CDM model with the same value of $\Omega_0$
(OCDM; $\Omega_0=0.4$, $h=0.5$, $n=1$). 
Though it appears that using $n_{R}$ in the $\varepsilon$ term in
equation~(\ref{S3}), with our nonlinear-bias model, is not strictly
correct for scale-dependent power spectra (see Figure \ref{fig:S3biasdeg}),
we can at least qualitatively infer from Figure \ref{fig:epscorrection}
that the contribution of the $\varepsilon$ term
to $S_3$ can become significant on
scales $R \ga 100$ $h^{-1}$ Mpc, particularly for models with higher
values of $\Gamma$. 
Thus, when considering data on the moments of counts in cells, smoothed on
these scales, this correction term must be included 
to properly
account for the presence of nonlinear bias. BKJ obtain
a similar correction to the angular skewness, and corresponding corrective
terms should arise in leading-order perturbative calculation of
higher-order moments, such as the kurtosis, $S_4$, and so on. 
For smoothed bias, as defined in 
equation~(\ref{dexpR}), this result implies that if $b_1^{(s)}$ and
$b_2^{(s)}$ are defined at some smoothing radius $R_1$, then at a
sufficiently large
scale $R_2 \neq R_1$, $b_2^{(s)}$ will acquire a different value
due to the $\varepsilon$ induced by the change in $R$. 
For a given CDM model, measurements of $S_3$ using different smoothing
radii should thus yield identical results for $b_1=b_1^{(s)}$ and $b_2$, 
but different
results for $b_2^{(s)}$.\footnote{We thank
Roman Scoccimarro for clarification of this point.} This suggests
that the linear and nonlinear bias terms may be distinguishable,
in a model-dependent fashion, 
on the basis of skewness measurements alone, provided however, 
that $S_3$ can
be measured over scales sufficiently larger than 100 $h^{-1}$ Mpc.  
An illustration of this point is shown below, in Figure \ref{fig:S3biasdeg}.
If, however, measurements do not show any disagreement between 
$b_2$ and $b_2^{(s)}$ on large scales, this would provide evidence
to support scenarios such as peak-biasing or halo-biasing models and
argue against biasing as defined in equation~(\ref{dexp}).


\begin{figure}[htbp]
\plotone{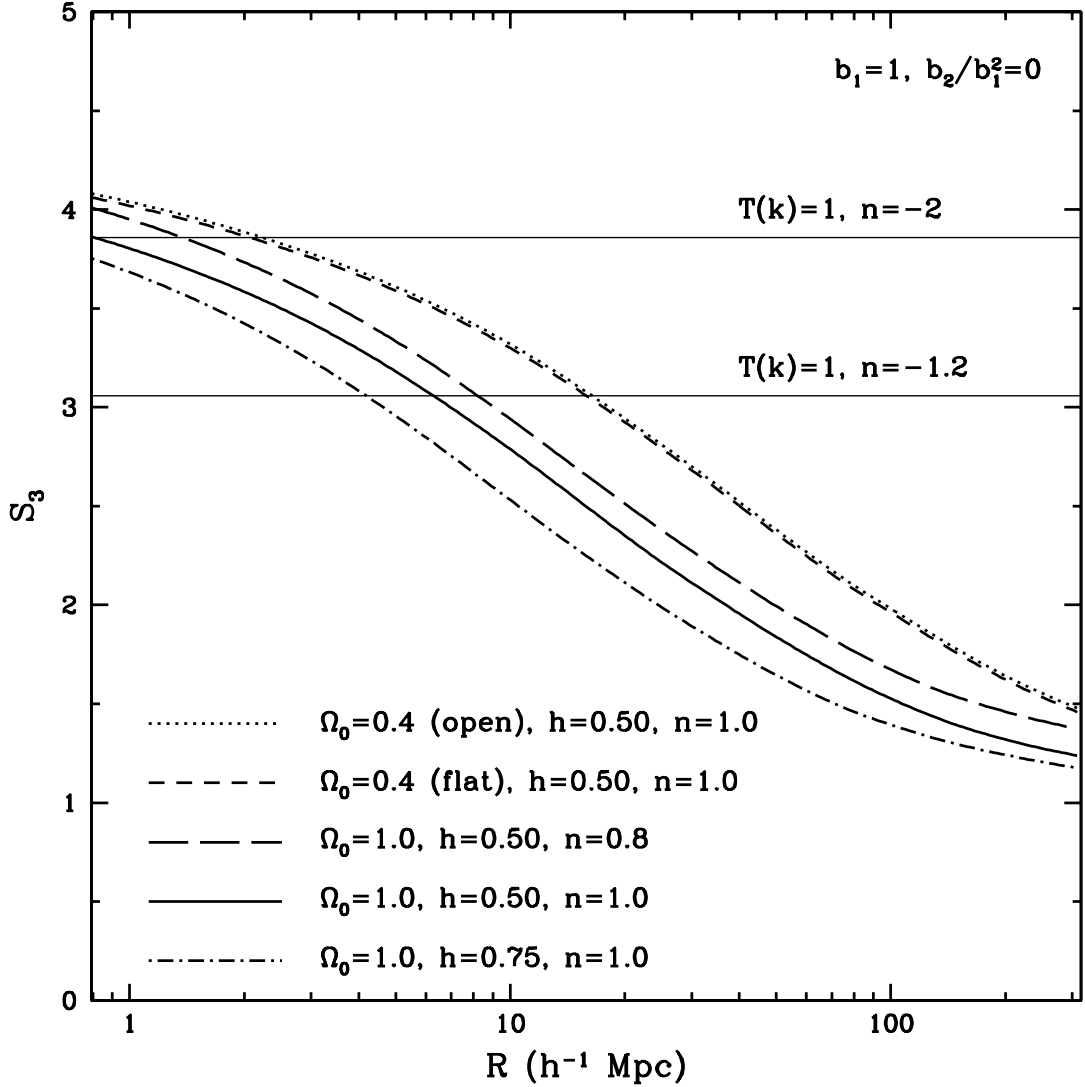}
\caption{The predicted normalized skewness, $S_3(R)$, for an unbiased 
tracer-mass
distribution smoothed with a spherical top-hat filter of radius $R$, for
the four flat and one open cosmological models described in the text.
The scale dependence of $S_3$ arises from the adopted CDM transfer function. 
By comparison, the 
thin horizontal lines are the results obtained using power-law
spectra with $n=-2$ and $n=-1.2$. 
Note that the dependence of $S_3$ on the assumed
cosmological model is fairly weak, particularly if the observed constraint,
$0.2 < \Omega_0 h < 0.3$, is imposed.}
\label{fig:S3}
\end{figure}

Since there is no known method of
obtaining analytic results for the general case of the full 3PCF, assuming
scale-dependent CDM power spectra, we will hereafter, for consistency,
only present results obtained by numerical integration, unless
otherwise noted.
Many of the
integrals throughout this paper involve highly oscillatory
integrands, making them difficult to evaluate. We have checked
that our numerical results for $S_3$ in particular
agree to high precision with those obtained using Bernardeau's (1994a)
analytic prescription, and that our results in general are reliable
to within a few percent.  
Figure \ref{fig:S3} displays our numerical results for $S_3(R)$
for an unbiased ($b_1 = 1$, $b_2 =0$) tracer-mass population in
four flat 
($\Omega_0+\Omega_\Lambda=1$) cosmological models:
SCDM, TCDM, $\Lambda$CDM (each from above), and CDM with a high
Hubble parameter (HCDM; $\Omega_0=1$, $h=0.75$, $n=1$), as well as 
the open CDM model (OCDM, from above).  

As expected, Figure \ref{fig:S3} shows that the CDM power
spectra introduce a dependence on the smoothing scale, $R$,
and [compared to equation~(\ref{S3})]
a more substantial dependence on the combination
of cosmological parameters in $\Gamma$.
Still, the different cosmological models
represented, though spanning a fairly broad range,
yield fairly similar results for $S_3$, certainly all
consistent within the typical errors of current observations 
(e.g., Gazta\~{n}aga 1992; Jing \& B\"{o}rner 1998). 
The slight variation seen is due primarily
to the dependence on $\Gamma$;
there is little difference between
the SCDM and TCDM models (which differ only in the value of $n$)
and virtually no difference between the OCDM and $\Lambda$CDM models,
as expected from the very weak $\mu$ dependence.
If the analysis is restricted to the
range $0.2 < \Gamma < 0.3$, as suggested by current data
(e.g., Bartlett \etal 1998), one can infer
from the Figure that the predicted values for $S_3$ are relatively
insensitive to the adopted cosmological model.    
In the $R \rightarrow 0$ limit, the untilted
($n=1$) CDM curves tend toward the expected value of
$(34/7)=(34/7)-(3+n_{k})$ obtained with $n_{k} \equiv 
[d \log{P(k)}]/d\log k=-3$,
the value of the effective spectral index of the (untilted) linear-theory CDM
power spectrum at small scales; for tilted models, $S_3$ tends
toward a value $n_k-1$ smaller.  At large $R$, they tend toward
$(6/7) = (34/7)-(3+n_{k})$, where $n_{k}=1$.
These perturbative results are expected to be valid
at scales $R\gtrsim 5-10$ $h^{-1}$ Mpc, although as mentioned
above, there are reasons to expect them to be reliable even at
smaller scales.
In practice, a comparison of these predictions with data will be limited
at small scales by increasingly significant nonlinear effects, 
as well as Poisson shot noise arising from the discreteness
of the counts-in-cells (Peebles 1980, Gazta\~{n}aga 1994), 
and at very large scales
by sampling noise arising from small number of independent cells on the sky.
Note for comparison, that the thin horizontal lines, which show the 
(virtually $\Omega_0$-independent) analytic results
for scale-free
power spectra with $n=-2$ and $n=-1.2$ [the canonical index value obtained
from measurements of the 2PCF at small scales (Peebles 1980)],
provide poor fits to the CDM predictions. 





\subsection{Effects of Bias and its Evolution}

The calculations above can be generalized by allowing the bias
parameters to evolve with time, as advocated by theory and
observations.  Peacock (1997) predicts that the linear 
galaxy-mass bias term evolves from a value of roughly 6 at a
galaxy-formation redshift of $z_f \simeq 6-8$, to a nearly
unbiased value of $\gtrsim 1$ today,  
with all models of structure formation requiring
some degree of bias at $z \gtrsim 3$. Matarrese \etal (1997)
explore several different models of bias evolution, and also
find that the bias for objects of between $10^{10}$ and
$10^{13}$ $M_{\odot}$ evolves from a value of a few at $z \simeq 5$
to roughly unity today. Studies of the clustering strength of tracer
populations that span an adequately large range of redshifts,
such as radio galaxies and Lyman-break galaxies, show
evidence for a large bias at high redshift, which decreases with
time in a manner consistent with the predictions above.
(Steidel \etal 1998; Cress \& Kamionkowski 1998).  Assuming that
objects form at a fixed redshift $z_f$ by some arbitrary local
process which induces a bias at that epoch, and are
subsequently governed purely by gravity, Fry (1996) derives
the result for the bispectrum of a tracer population in an
Einstein-de Sitter universe, including the time-dependence
of the bias parameters.  Matarrese \etal (1997), refer to
this as the ``object-conserving'' model,  since it does not
account for merging (see below). Tegmark \&
Peebles (1998) generalize the Fry model to the case where the
dimensionless mass-galaxy correlation coefficient may be
different than unity, but we take the Fry model as a reasonable
starting point for the exploration of the effects of bias
evolution.

It is straightforward to generalize the bias-evolution model of
Fry (1996) to an arbitrary expansion history.  Doing so, we
obtain for the bispectrum in these models,
\begin{eqnarray}
     B({k}_{1},{k}_{2},{k}_{3},t) & = & P(k_{1},t)P(k_{2},t)
     \left[ C_1(t)
     + C_2(t) \cos\theta \left(\frac{k_{1}}{k_{2}}+ 
     \frac{k_{2}}{k_{1}}\right) 
     + C_3(t) \cos^2 
     \theta \right] \nonumber \\
     &   & + \mbox{2 permutations},
\label{Bfe}
\end{eqnarray}
where
\begin{eqnarray}
     C_1(t) & = & \frac{(10/7)d^2(t) + 2(b_{1_{*}}-1)[d(t)-2/7] + b_{2_{*}}}
     {[d(t)+b_{1_{*}}-1]^2}, \\
     C_2(t) & = & \frac{d(t)}{d(t)+b_{1_{*}}-1}, \\
     C_3(t) & = & \frac{(4/7)[d^2(t) + b_{1_{*}} -1]}{[d(t)+b_{1_{*}}-1]^2},
\label{Cs}
\end{eqnarray}
$d(t)=D(t)/D(t_{*})$, and a subscripted asterisk denotes
the value of that parameter at the epoch of formation. 
Note that we have now ignored the very weak dependence of the bispectrum
on $\mu$; the dependence on the expansion history, i.e., on
the species contributing to the total energy density, is contained
in the growth factor, $D(t)$. 

In this model, the linear-bias term 
effectively evolves as
\begin{equation}
     b_1(t) = \frac{d(t) + b_{1_{*}} -1}{d(t)},
\label{b1e}
\end{equation}
decaying towards unity, as in most bias models, as the observed matter
settles into the potential wells of the underlying distribution.
We note that if, in reality, merging of the tracer-mass population does in fact
play a considerable role, this could result in an anti-bias ($b_1 < 1$).
Anti-bias has been seen in some $N$-body simulations (Jenkins \etal 1997), 
and certain
models of structure formation do require some degree of anti-bias to
reconcile with current observations.
The above model for bias evolution, however, will never yield $b_1(t) < 1$
if $b_{1_{*}} >1$. To explore this possibility, one might wish to
employ other bias-evolution models which account for merging. Here, however,
we are interested primarily in the qualitative dependence of the results
on the dominant form of bias variation, namely, its decay with time from
initially large values at high redshift.

The Einstein-de Sitter result
for $S_3$ with bias evolution, assuming a scale-free $P(k)\propto k^n$ is
presented in Fry (1996).  For arbitrary $E(z)$, the result becomes
\begin{equation}
     S_3 = \frac{d^2(t)(34/7) + (b_{1_{*}}-1)[6d(t) - 8/7] + 3b_{2_{*}}}
     {[d(t)+b_{1_{*}}-1]^2} - (3+n)\frac{d(t)}{d(t)+b_{1_{*}}-1}.
\end{equation}
Here, we extend the calculation 
to the case of true, scale-dependent CDM power spectra, 
by using the bispectrum
of equation~(\ref{Bfe}) in (\ref{skewness}), which amounts to making the 
substitutions
\begin{eqnarray}
     \frac{1}{b_1} \longrightarrow C_2(t), \:\:\:\:
     \frac{1-\mu}{b_1} \longrightarrow C_3(t), \:\:\:\:
     \frac{b_2}{b_1^2} \longrightarrow C_1(t) -2C_2(t) + C_3(t),
\label{subs1}
\end{eqnarray}
in equation~(\ref{skewness2}).

\begin{figure}[htbp]
\plotone{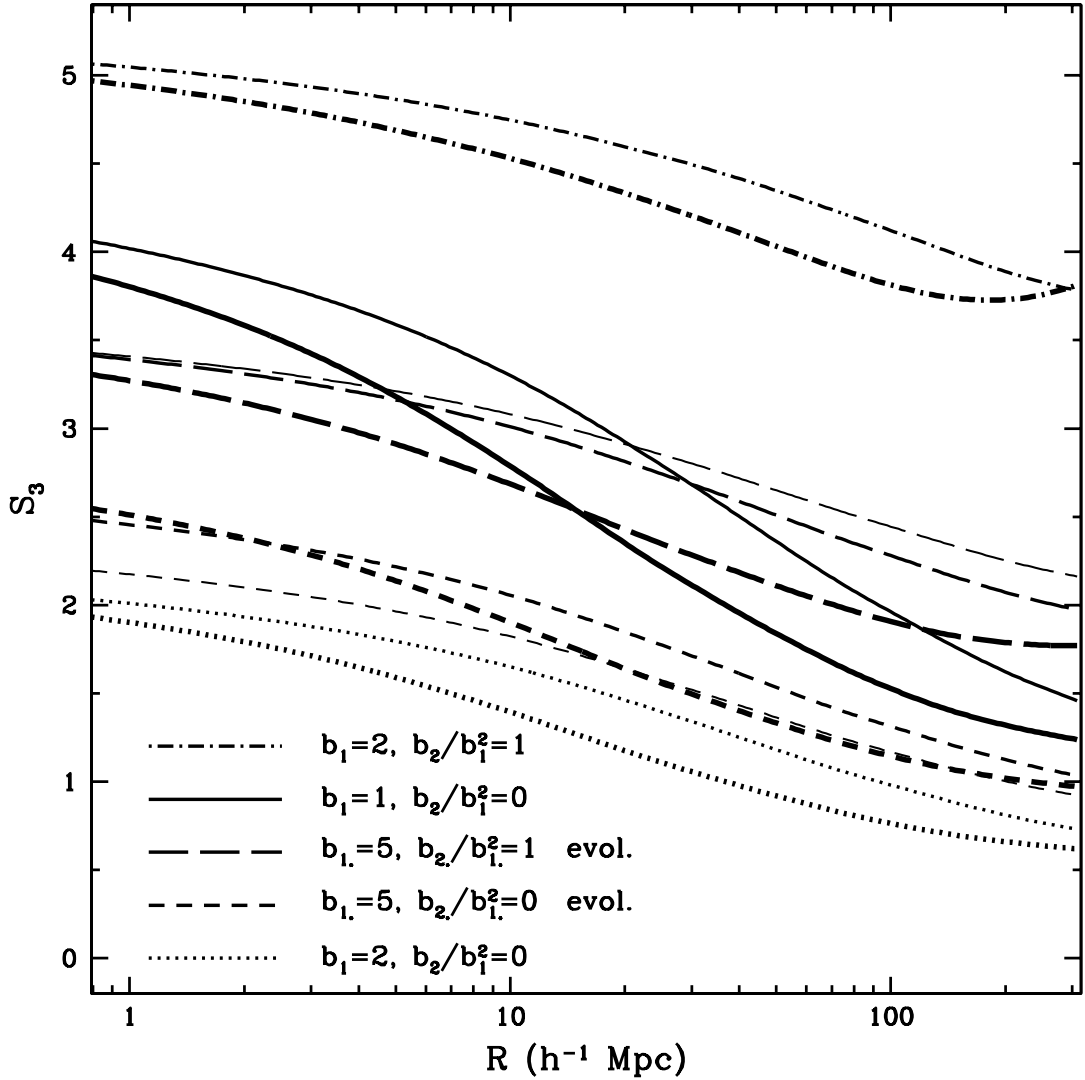}
\caption{The predicted normalized skewness for the five different bias
scenarios described in the text. For each case, the SCDM, $\Lambda$CDM,
and OCDM models are shown in thick, medium, and thin lines, respectively
(for the non-evolving bias scenarios, the latter two models are identical).
Note that the dependence on the overall biasing scheme far outweighs that
on the cosmological parameters within a given scheme. In particular,
a linear(nonlinear) bias term lowers(raises) the predicted curve with 
respect to
the unbiased case, and a significant linear bias term flattens the curve
appreciably.}
\label{fig:S3bias}
\end{figure}

Figure \ref{fig:S3bias} displays the results for five different 
bias scenarios, each
employing the SCDM, $\Lambda$CDM, and OCDM models above:  
an unbiased scenario 
($b_1=1$, $b_2/b_1^2=0$, taken from Figure 2), 
a non-evolving, linear bias ($b_1=2$, $b_2/b_1^2=0$),
a non-evolving, nonlinear bias ($b_1=2$, $b_2/b_1^2=1$),
an evolving, linear bias ($b_{1_{*}}=5$, $b_{2_{*}}/b_{1_{*}}^2=0$),
and an evolving, nonlinear bias ($b_{1_{*}}=5$, $b_{2_{*}}/b_{1_{*}}^2=1$).
For the non-evolving bias scenarios, the $\Lambda$CDM and OCDM models
are taken to be identical, as suggested by Figure 2. These models do,
however, differ in the case of evolving bias due to the different
time dependence of the linear growth factor, and the differences 
between them would increase with decreasing $\Omega_0$.
For the evolving cases we assume a formation redshift of $z_f = 5$. 
In calculating the expressions involving $b_2$, we must evaluate
sums of the kind in equation~(\ref{skewness2}). 
Numerical tests show that these converge 
rapidly, typically
requiring at most three terms to achieve percent-level accuracy.

These models are intended as an illustrative, but by no means
exhaustive, representation of the possibilities.  Even so,
it is clear from Figure \ref{fig:S3bias} that
that $S_3(R)$ depends far more sensitively on the
overall form of the bias than on the values of $\Gamma$ or $\Omega_0$
within an individual biasing scheme. 
For example, adding a very slight linear bias to the low-$\Gamma$
model would yield a result similar to the unbiased SCDM prediction, within
current error estimates. Generally, the presence of any significant linear
bias,
$b_1 > 1$, flattens the $S_3(R)$ curve, as compared with the unbiased case
(solid lines).  Furthermore,
an observed $S_3(R)$ curve which is well below the predicted unbiased result
can only be achieved by a significant linear bias term (assuming $b_2 \geq 0$),
while one well above this value can only arise from the existence of a 
non-zero $b_2$ term or from anti-biasing ($b_1 < 1$).
In general, the curves for an evolving bias produce less drastic shifts
than their non-evolving counterparts, despite the fact that the respective
bias terms are initially larger.  This is because the evolution
towards an unbiased
state effectively mimics a smaller, constant bias [c.f. equation~(\ref{b1e})]. 
Note that the SCDM models with nonlinear bias exhibit an upturn at large
$R$, arising from the contribution of the higher-order terms
in the sum in equation~(\ref{skewness2}). The other CDM models with 
nonlinear bias show an upturn only at larger scales, as suggested by 
Figure \ref{fig:epscorrection}.

\begin{figure}[htbp]
\plotone{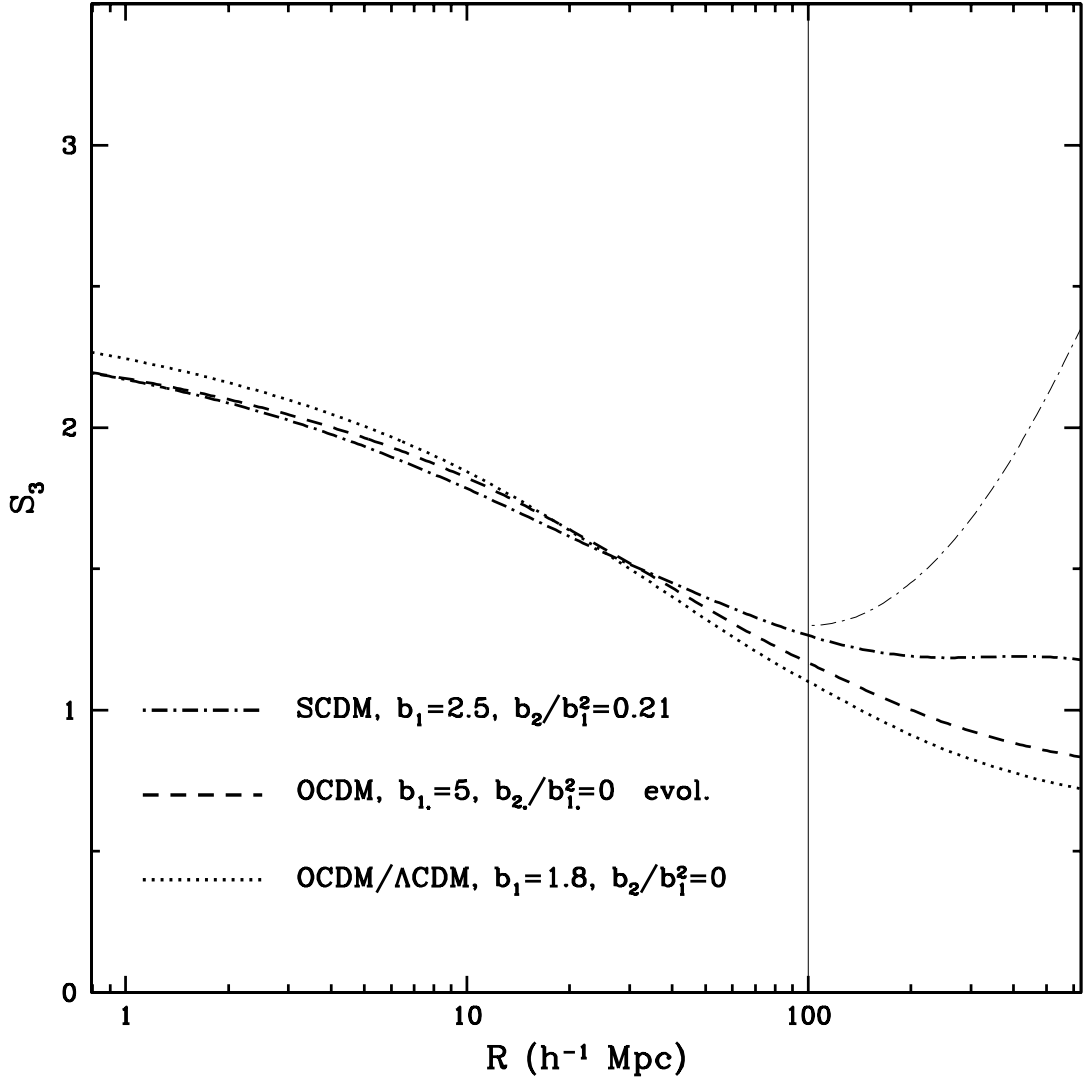}
\caption{An example of nearly degenerate
predictions for $S_3(R)$, arising from the fact that the skewness
provides only one constraint on the combination
of the cosmological and bias parameters, and the evolution of
the latter. In this case, the SCDM model with constant nonlinear bias, 
OCDM model with evolving linear bias, and
OCDM/$\Lambda$CDM model with constant linear bias are nearly identical
out to smoothing scales of $R \simeq 100$ $h^{-1}$ Mpc. Above this scale,
the SCDM model with nonlinear bias can be better distinguished due
to the different scale dependence introduced by higher-order terms
in the sum in equation~(\ref{skewness2}). For comparison,
the thin dot-dashed line
shows the SCDM result using equation~(\ref{S3}) with $\varepsilon[n_{R}(R)]$,
and appears to overestimate the impact of the nonlinear term
at large scales.}
\label{fig:S3biasdeg}
\end{figure}

Figure \ref{fig:S3bias} also reveals that, despite the separate
dependences of $S_3$ on $b_1$, $b_2$, and their evolution, different biasing
scenarios may be
very difficult to distinguish on the basis of the skewness 
measurements alone.
Different sets of parameter choices could easily yield
curves for the various scenarios which essentially overlap, requiring
high-accuracy data to differentiate them.
An example of this is shown in Figure \ref{fig:S3biasdeg}, where we
see that, for scales $R < 100$ $h^{-1}$ Mpc,
the results from an SCDM model 
with constant, nonlinear bias, an OCDM model with evolving,
linear bias, and an OCDM/$\Lambda$CDM model with constant, linear bias,
are very similar.
The OCDM model with evolving, linear bias is taken from Figure
\ref{fig:S3bias}, and according to equation~(\ref{b1e}), is
expected to be similar to an open model with a constant value
of $b_1 = 2.04$, or a $\Lambda$CDM model with a constant $b_1 = 1.8$,
and indeed Figure \ref{fig:S3biasdeg} shows this to be the case.
For an inferior data set, this degeneracy
between the cosmological parameters, and in particular $b_1$,
$b_2$, and their time dependence, cannot be broken with the
skewness, measured on scales
$R < 100$ $h^{-1}$ Mpc,
as it provides only one constraint on the
combination of these quantities. 
Note, however, that for $R \ga 100$ $h^{-1}$ Mpc, the SCDM 
model with nonlinear bias can be
better distinguished due to the different scale dependence arising from the
higher-order terms in the sum in equation~(\ref{skewness2})
For comparison,
the thin dot-dashed line shows the same
result obtained using $\varepsilon[n_{R}(R)]$ in equation~(\ref{S3}), 
which appears to
overestimate the numerical result (thick lines) obtained by evaluating the
the sum in equation~(\ref{skewness2}). If this large-scale variation induced
by nonlinear biasing is observed, it would allow better constraints
on both the linear and nonlinear 
bias parameters solely from measurements of
$S_3$. The absence of this behavior would still provide
valuable insight, arguing against our bias model
and in favor of smoothed-bias models, where $b_1$ and $b_2$ tend towards
constants at large scales. 
However,
it is at present unlikely that $S_3$ could be measured on such
large scales 
with sufficient precision to address these issues.

\section{SPATIAL THREE-POINT CORRELATION FUNCTION}

The one-point statistic $S_3$, is a volume-averaged 
quantity and thus, while it is seen to preserve information about the overall
dependence of clustering strength with scale, discards detailed
information about the configuration dependence contained in the full
3PCF.  It has been shown, for example, that the configuration dependence
of the bispectrum and 3PCF
in leading-order PT can be used to separate the contributions
of gravitational clustering and bias to the observed power
(FG93;
Fry 1994; Frieman \& Gazta\~{n}aga 1994),
as well as obtain multiple, independent constraints on the combination
of the constant $b_1$ and $b_2$ parameters
(Jing \& Zhang 1989; Gazta\~{n}aga \& Frieman 1994;
Fry 1994, 1996; Jing 1997; Matarrese, Verde, \&
Heavens 1997).  

Much attention has therefore been focused
on the normalized spatial 3PCF,
\begin{equation}
Q({r}_{12},{r}_{23},{r}_{31}) = \frac{\zeta({r}_{12},{r}_{23},{r}_{31})}
{\xi(r_{12})\xi(r_{23}) + \xi(r_{12})\xi(r_{31}) + \xi(r_{23})\xi(r_{31})},
\label{Qdef}
\end{equation} 
where
the lengths $r_{ij}$ form a triangle in real space. We can evaluate this
expression using
equations~(\ref{xi}) and (\ref{zetaR}), where we no longer include the window 
functions since we presume that the CFs are obtained from
direct counting, rather than from smoothed counts in cells. 
In this case, it is more natural to use the unsmoothed bias parameters
defined by equation~(\ref{dexp}), rather than the smoothed bias
parameters of equation~(\ref{dexpR}). Using
the techniques and conventions
of the previous sections, we obtain, after some manipulations, the spatial 
2PCF,\footnote{Had we wished to calculate the 2PCF using counts in cells, 
there would be an additional factor of $W^2(xR/r)$ in the integrand, 
and corresponding terms in the 3PCF, which would alter the results on scales
where $R/r$ is non-negligible (i.e., separations comparable to
the cell size), to account for the effects of smoothing on correlations 
at these scales. The dependence of the CFs on the ratio between the 
smoothing half-width and the pair separation can be appreciable;
for scale-free power spectra with $n=-2.2,-1.2,-0.2,0.8$ this
term adds corrections to $\xi$ of 1\%, 10\%, 49\%, and
294\% for $r=2R$, dropping to 0.14\%, 1.2\%, 4.3\%, and 2.9\% for
$r=5R$. For a CDM power spectrum, with $\Gamma = 0.5$, 
the corrections are more significant:
42\%, 11\%, and 2.4\% for $r/R$ values of 2, 5, and 10, respectively.
For counts-in-cells CFs, one must take care to restrict
measurements to scales significantly larger than the smoothing radius,
or include the smoothing terms in the theoretical calculations.}
\begin{equation}
    \xi (r) = \frac{1}{2\pi^2} \int dk \: k^2 P(k) \frac{\sin{kr}}{kr}
    = \frac{Ab_1^2D^2(t)}{2\pi^2 r^{n+3}} \int dx \: x^{n+2} T^2(x/r) 
    \frac{\sin{x}}{x},
\label{2full}
\end{equation}
and 3PCF,
\begin{figure}
\plotone{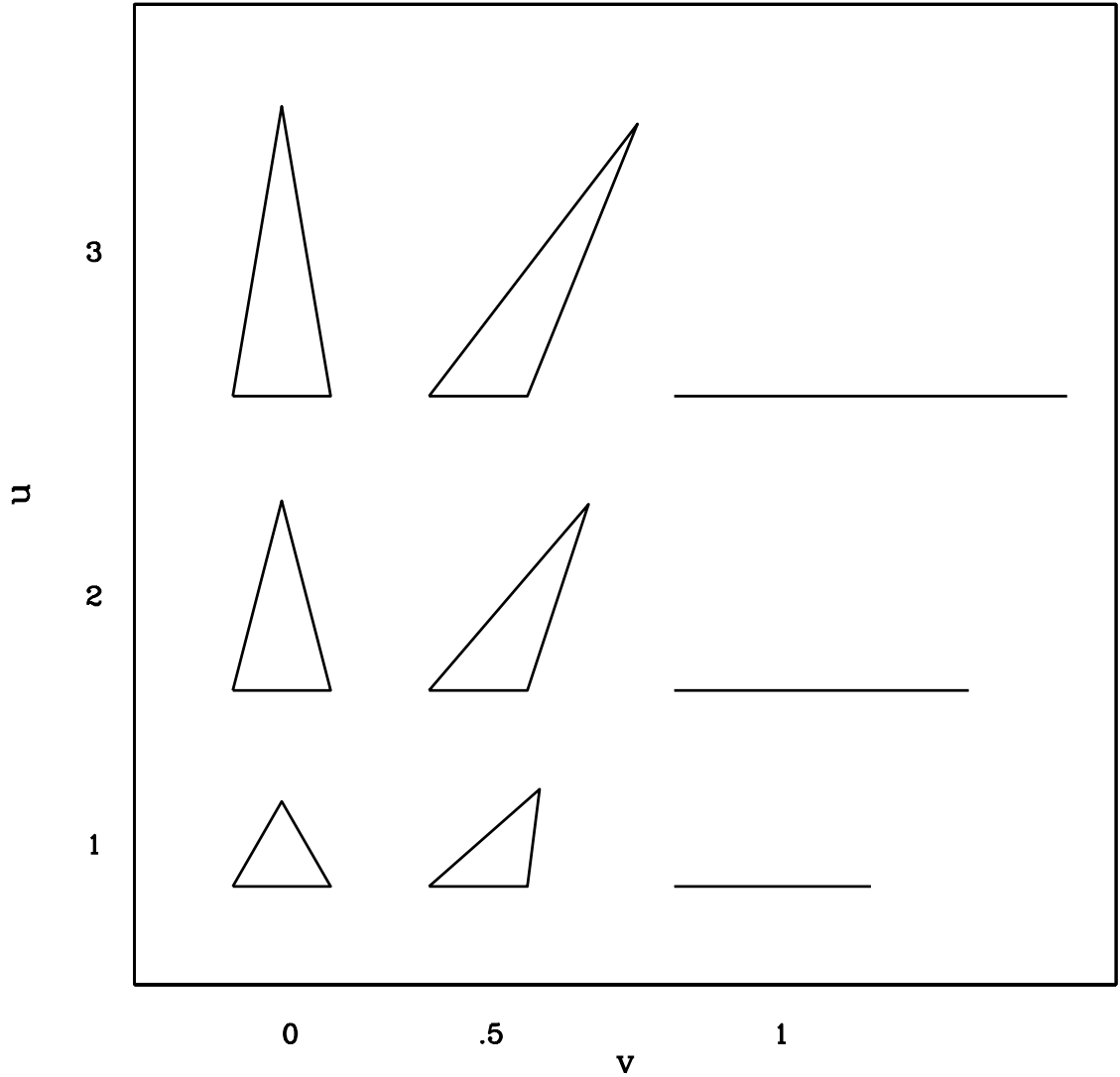}
\caption{The variation of triangle geometry with $u$ and $v$, for a fixed
value of $r$ given by the side of the $u=1$, $v=0$ equilateral triangle.
Triangles with $u=1$ are isosceles with the shorter sides equal,
those with $v=0$ are isosceles with the longer sides equal, and
$v=1$ yields a colinear configuration. Larger values of $u$ and/or $v$
pick out more elongated structures.}
\label{fig:tri}
\end{figure}
\begin{eqnarray}
     \zeta(r_{12},r_{23},r_{31}) & = & \frac{A^2 b_1^4 D^4(t)}{4\pi^4}
     \frac{1}{r_{31}^{n+3}r_{23}^{n+3}}
     \int_0^\infty dx_1 x_1^{n+2} T^2(x_1/r_{31}) 
     \int_0^\infty dx_2 x_2^{n+2} T^2(x_2/r_{23}) \nonumber \\
     &   & \times \Biggl\{ \left( \frac{1+\mu}{b_1} + \frac{b_2}{b_1^2} \right)
     \frac{\sin{x_1}}{x_1}\frac{\sin{x_2}}{x_2}  \nonumber \\ 
     &   & - \frac{1}{b_1}\cos{\theta_{12}}
     \left( \frac{r_{23}}{r_{31}}  x_1^2 V(x_1)V(x_2)
      + \frac{r_{31}}{r_{23}} 
      x_2^2 V(x_1)  V(x_2) \right) \nonumber \\
     &   & +\frac{1-\mu}{b_1} \biggl[ \cos^2{\theta_{12}}  Y(x_1) Y(x_2)
     + 3   V(x_1)  V(x_2)   \nonumber \\
     &   & -  Y(x_1) V(x_2)
     - Y(x_2) V(x_1)   \biggr] \Biggr\} + \mbox{(cyc.)},
\label{zeta2}
\end{eqnarray}
where $\cos\theta_{12}$ is the angle between $\bfr_{13}$ and $\bfr_{23}$,
$V(x) = W(x)/3$, and
and $Y(x) \equiv W(x)- (\sin x)/x$. For 
an evolving bias, we simply change 
\begin{equation}
     \left( \frac{1+\mu}{b_1} + \frac{b_2}{b_1^2} \right)  \longrightarrow  
     C_1(t), \:\:\:\:
     \frac{1}{b_1}  \longrightarrow  C_2(t), \:\:\:\: 
     \frac{1-\mu}{b_1} \longrightarrow  C_3(t).
\end{equation} 

Equation~(\ref{zeta2}) does not contain terms analogous to the
higher-order terms in the sum in equation~(\ref{skewness2}) [or to the
$\varepsilon(n)$ term in equation~(\ref{S3})], since evaluation
of the 3PCF by direct counting requires no smoothing. With our bias model, 
such terms would be induced, however, from the calculation of the 3PCF using
counts in cells (see footnote). One must be careful to realize that
the nonlinear-bias parameter measured from the direct-counting 3PCF
in equation~(\ref{zeta2}) will be different from the smoothed
value inferred from the FG93 result for $S_3$ (or, equivalently,
from that inferred from the counts-in-cells 3PCF), though the
two can be related, e.g., via the $\varepsilon(n)$ term in
equation~(\ref{S3}).

The dependence of $\xi(r)$ on $r^{-(n+3)}$ in the prefactor of
equation~(\ref{2full}) is the familiar result, yielding
$\xi(r) \propto r^{-1.8}$ for a scale-free power spectrum
with the canonical value of $n= -1.2$ (Peebles 1980). 
For more realistic CDM power spectra, there
is an additional dependence on $r$ arising from the scale dependence
of the transfer function.  Similarly, had we ignored the shape-dependent
terms in the bispectrum [equation~(\ref{Bf2})], we would arrive
at the standard empirical ``approximation'' for the 3PCF, 
$\zeta(r_{12},r_{23},r_{31}) \propto 
\xi(r_{12})\xi(r_{23}) + \xi(r_{12})\xi(r_{31}) + \xi(r_{23})\xi(r_{31})$. 
However, the 
true scale and shape dependences of the 3PCF, as manifested in 
equation~(\ref{zeta2}), can be exploited to gain far more information
(Fry 1984; Jing \& B\"{o}rner 1997).  Following
Peebles \& Groth (1975), we take
\begin{equation}
    r=r_{12}, \:\:\:\:\:\: u=\frac{r_{23}}{r_{12}}, 
     \:\:\:\:\:\: v=\frac{r_{31}-r_{23}}{r_{12}}
\end{equation}
as the defining parameters for a triangle, where $r_{12}<r_{23}<r_{31}$,
so that $u \ge 1$ and $0 \le v \le 1$. For this convenient choice
of parameters, $r$ fixes the overall size of the triangle, while
$u$ and $v$ determine the exact size and shape. Figure \ref{fig:tri} shows the
variation of the geometry with $u$ and $v$ for a given choice of $r$.

\begin{figure}[htbp]
\plotone{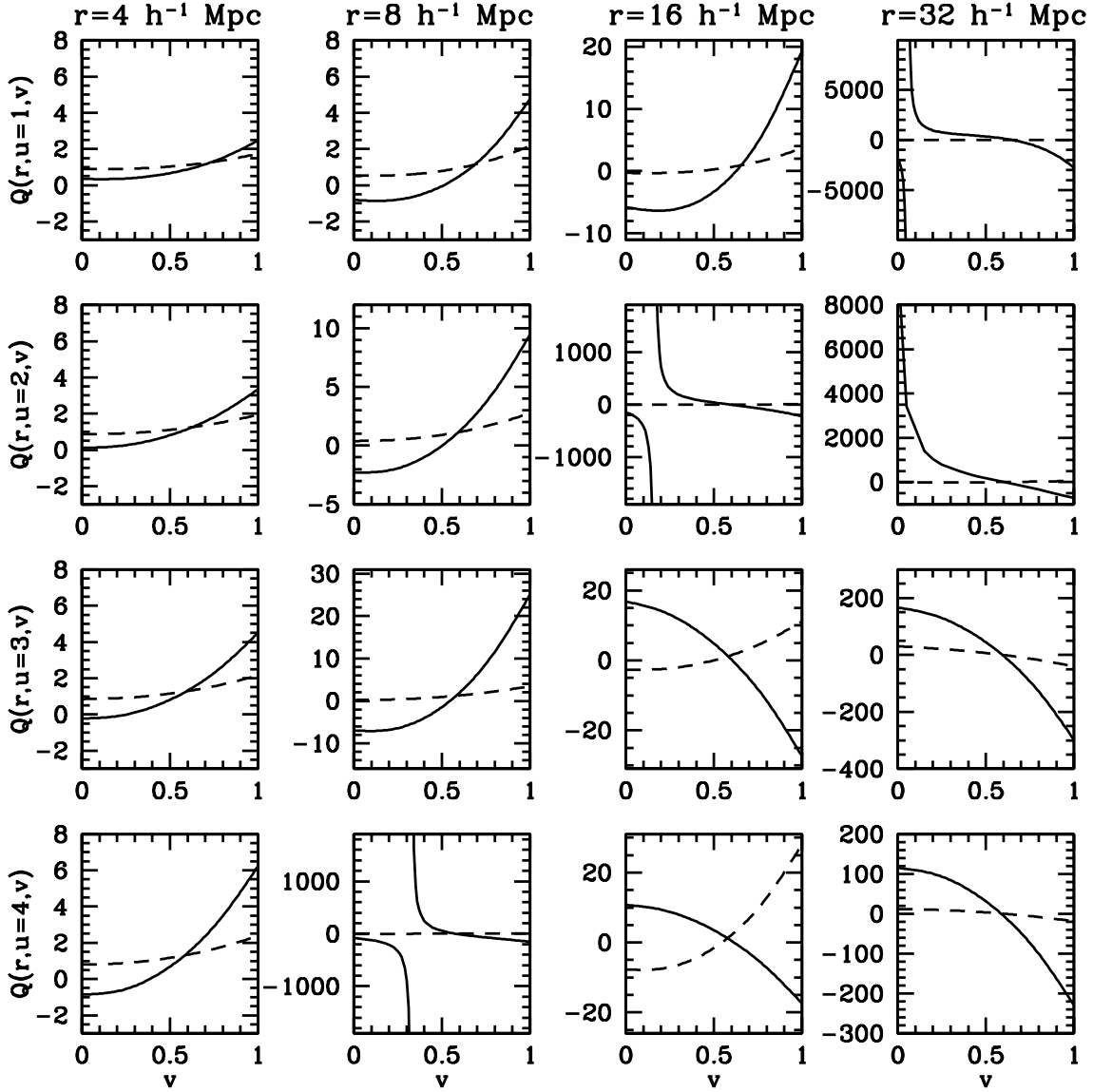}
\caption{Predicted normalized 3PCF, $Q(r,u,v)$, for the $\Gamma=0.5$ (SCDM,
solid lines)
and $\Gamma=0.2$ (OCDM/$\Lambda$CDM, dashed lines)
models without bias. The columns represent values of $r = 4$, 8, 16, and 32 $h^{-1}$ Mpc
(left to right) while the rows represent values of $u=1$, 2, 3, and 4
(top to bottom).  Note the different scalings on
the vertical axes. Over these scales, the SCDM model shows 
dramatic variation of $Q$
with triangle geometry, exhibiting divergences
in some cases.}
\label{fig:Q}
\end{figure}

\subsection{Results with no Biasing}

\subsubsection{The Problem With $Q$}

\begin{figure}[htbp]
\plotone{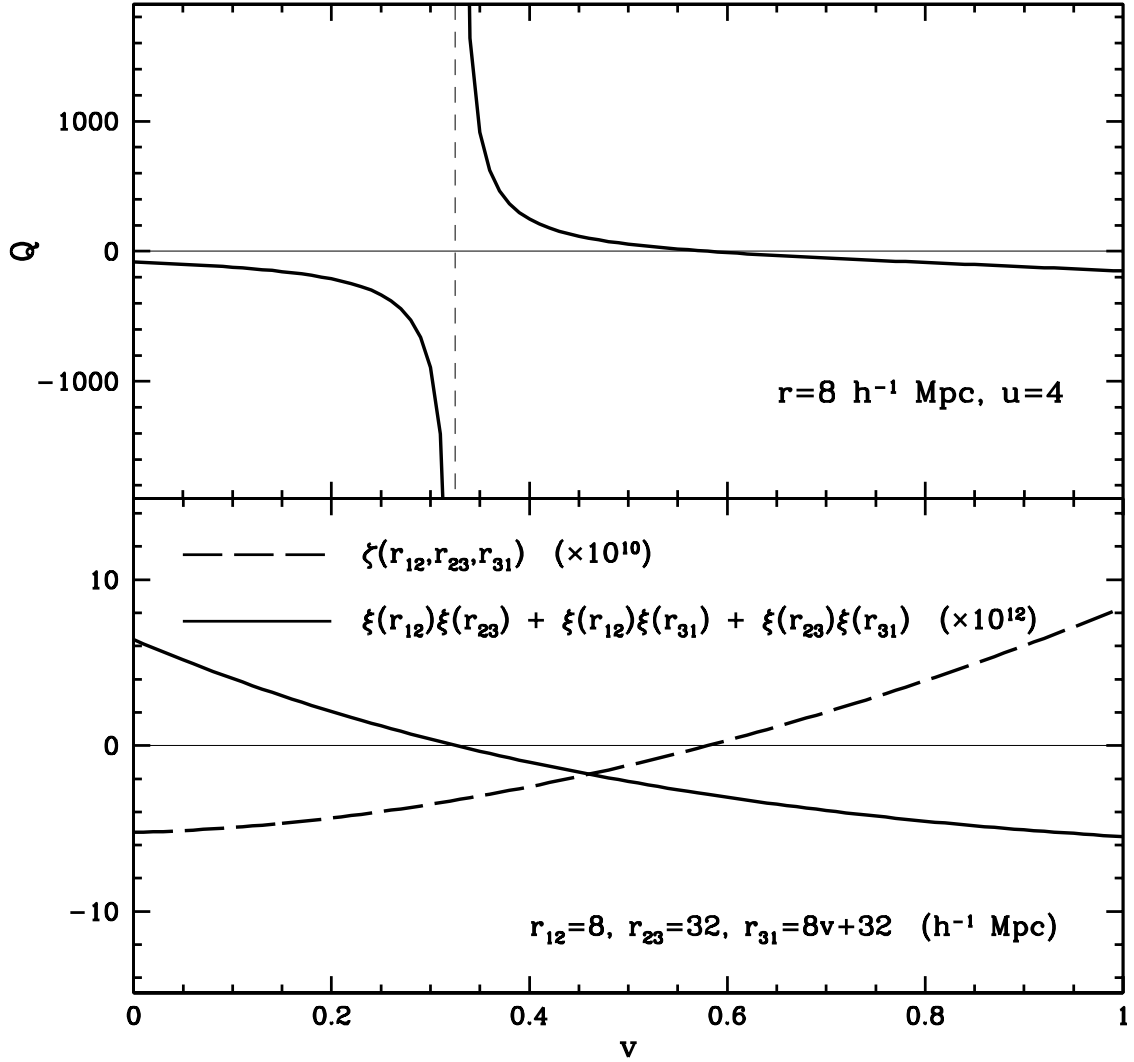}
\caption{The top panel shows the predicted $Q(v)$ for 
$r=8$ $h^{-1}$ Mpc and $u=4$, while the bottom panel shows the
separate behaviors of the numerator (multiplied by $10^{10}$; 
dashed line) and denominator
(multiplied by $10^{12}$; solid line) 
of equation~(\ref{Qdef}) for this configuration. The 3PCF itself is smoothly
varying; the large values of $Q$ arise from the small values of the
denominator, which passes through zero at $v \simeq 0.325$.}
\label{fig:Qexplode}
\end{figure}

Figure \ref{fig:Q} shows the normalized
spatial 3PCF, $Q$, plotted
as a function of $r$, $u$, and $v$, using the transfer function
of equation~(\ref{Tk}) for the $\Gamma=0.5$ (SCDM) and
$\Gamma=0.2$ (OCDM/$\Lambda$CDM) models, assuming no bias.
The weak $\mu$ dependence makes the OCDM and
$\Lambda$CDM models nearly indistinguishable on the scale of
Figure \ref{fig:Q}, so only the former is plotted. Note the 
different scalings and the dramatic, non-monotonic 
variation and divergences seen
for certain configurations, particularly in the SCDM model.
JB97 also derive QL PT predictions for $Q$ in CDM models and
find very similar results for its configuration dependence,
though they do not show cases where the behavior is
particularly egregious.
Comparing these results with those from $N$-body simulations, however,
JB97 and Matsubara \& Suto (1994) find that the
predicted differences between various CDM models in $N$-body simulations
are markedly less than those predicted by QL PT, and moreover that in
simulations, these models in general show far less rapid variation of $Q$
with geometry than is seen in the QL PT predictions. JB97
claim that this discrepancy is due primarily to the absence of
nonlinear corrections in the PT calculation, but we suggest
another possibility. In the top panel of Figure \ref{fig:Qexplode} we
show our prediction for $Q(v)$ in a $\Gamma=0.5$ model for
$r=8$ $h^{-1}$ Mpc and $u=4$. The lower panel reveals that the 
divergence in $Q$ is caused not by any irregular behavior
of $\zeta$ (which is seen to vary in a smooth, monotonic fashion), but
rather by the behavior of the denominator in equation~(\ref{Qdef}),
which vanishes in this case at $v \simeq 0.325$ (note that we
have multiplied the numerator and denominator of equation~(\ref{Qdef})
by $10^{10}$ and $10^{12}$, respectively, for clarity). In this sense,
$Q$ is a poorly defined quantity, whose variation may not reflect
any rapid variations in $\zeta$, but rather stems
from the manner in which $\zeta$ is normalized. For the cases
shown in Figure \ref{fig:Q}, $Q$ is particularly rapidly varying 
in the SCDM model only because the
region of the power spectrum effectively sampled by a value of $\Gamma=0.5$
happens to yield vanishing behavior of 
$\xi(r_{12})\xi(r_{23}) + \xi(r_{12})\xi(r_{31}) + \xi(r_{23})\xi(r_{31})$
for some of the chosen configurations; similar behavior would be seen
in other models at appropriately different scales.

We speculate that the above-mentioned 
disagreement with $N$-body predictions, which
show far less variation in $Q$,
may simply be due to the practical
inability to measure 
$\xi(r_{12})\xi(r_{23}) + \xi(r_{12})\xi(r_{31}) + \xi(r_{23})\xi(r_{31})$
to high precision near zero, thus
smoothing out the $N$-body result for $Q$. This conclusion is supported
by the fact that all previous investigations find that the QL PT
predictions for $S_3$, which is normalized by a positive-definite quantity
(i.e., the square of the variance), {\em do} agree with $N$-body results
over the scales of interest\footnote{Hui \& Gazta\~{n}aga (1998) 
note a separate measurement bias,
applicable to $S_3$ as well, arising from the fact
that the ratio of two unbiased estimators is not unbiased estimator.}
(Bouchet, Schaeffer, \& Davis
1991; Fry, Melott, \& Shandarin 1993, 
1995; Bernardeau 1994a; Baugh, Gazta\~{n}aga, \& Efstathiou 1995;
Colombi, Bouchet, \& Hernquist 1996; JB97). We propose
that this explanation for the discrepancy in $Q$ can be evaluated
against the nonlinear-effects argument of JB97 by considering $N$-body
results at earlier epochs, where nonlinearities in the distribution
of fluctuations are known to be small. We point out that the Fourier-space
${ \cal Q}$, defined in equation~(\ref{B}), is normalized by a 
positive-definite quantity, and does not suffer from this problem entailed
in the definition of the real-space $Q$.

\begin{figure}[htbp]
\plotone{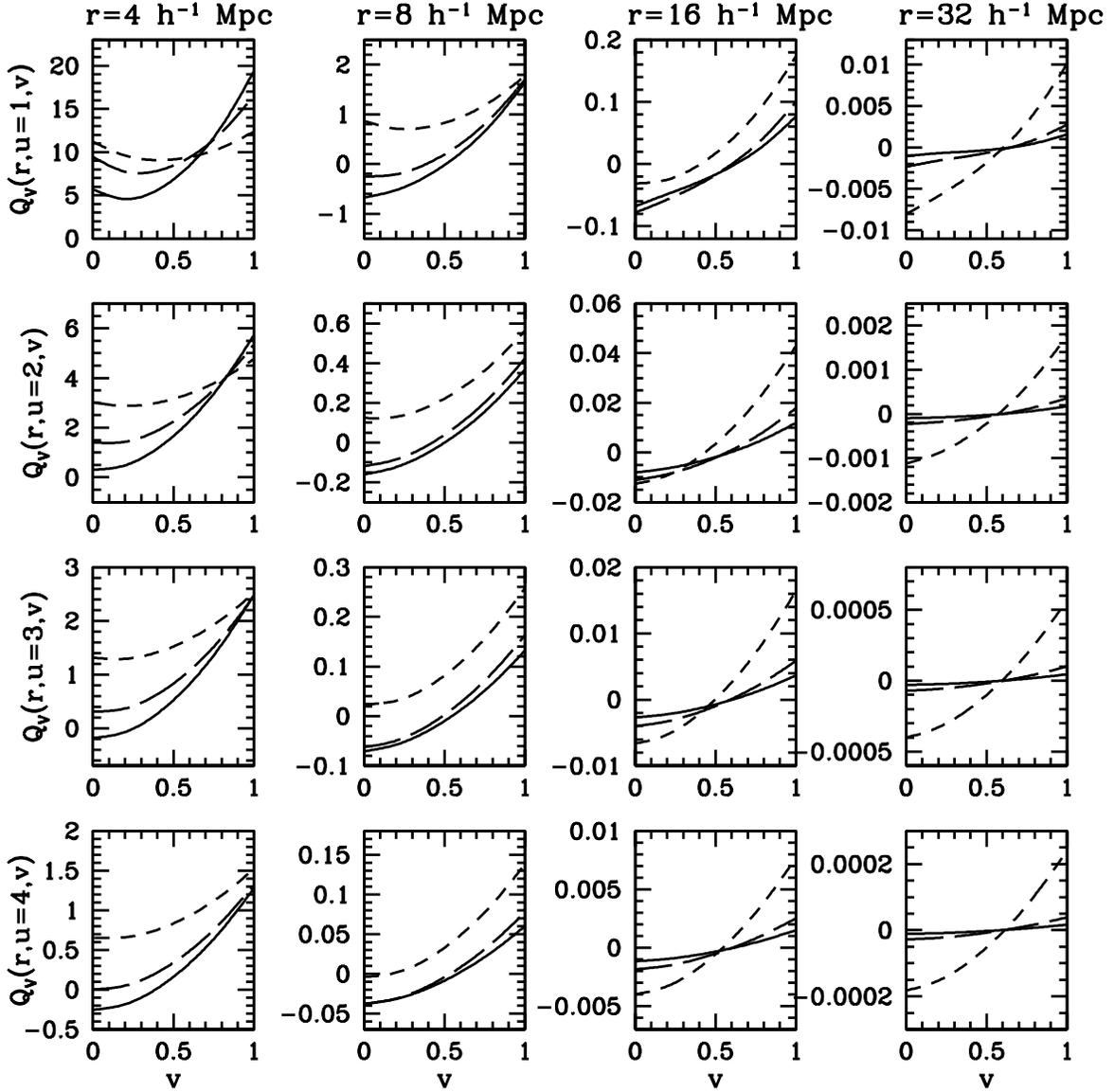}
\caption{Predicted $Q_V(r,u,v)$ for the unbiased SCDM
(solid lines), OCDM/$\Lambda$CDM (short-dashed lines), and
TCDM (long-dashed lines) 
models.
The columns represent values of $r = 4$, 8, 16, and 32 $h^{-1}$ Mpc
(left to right) while the rows represent values of $u=1$, 2, 3, and 4
(top to bottom).  Note the different scalings on the vertical
axes. All three models indicate 
that clustering strength decreases with scale
and increases with increasing $v$. Models with different
values of $\Gamma$, however, yield distinct predictions
for $Q_V$, though the SCDM and TCDM models, which differ only in the value
of $n$, are not well distinguished on large scales.}
\label{fig:zeta_actual}
\end{figure}

\subsubsection{Results for the 3PCF}

To avoid the above difficulties altogether, we define
$Q_V = \zeta/[\xi_{8 h^{-1} {\rm Mpc}}(0)]^2$; i.e., we normalize
the 3PCF (like $s_3$)
by the square of the variance at a fixed smoothing scale,
calculated for each model using equation~(\ref{xizerofinal}).
Thus, $Q_V$ is still independent of the power-spectrum
normalization, but does not exhibit the rapidly-varying or divergent
behavior seen
with $Q$, since we are now dividing by a positive-definite quantity.
The adopted smoothing scale is arbitrary; $8 h^{-1} {\rm Mpc}$ is
chosen merely for convenience.

Figure \ref{fig:zeta_actual} illustrates
our predictions for $Q_V(r,u,v)$
for the SCDM, OCDM/$\Lambda$CDM, and TCDM 
models with no bias.
Note that $Q_V$ is everywhere a smoothly-varying and
well-behaved function. Since, in each model, $Q_V$ is normalized by a
fixed quantity, the Figure clearly reveals the expected decrease in the
3PCF with increasing $r$. This behavior, however, is not
simply the $r^{-2(n+3)}$ variation predicted by the
hierarchical model, which also fails to produce the shape dependence
of the 3PCF. In general, we see that,  
within a given cosmological model, $Q_V$ increases
with increasing $v$.
Physically, this implies that clustering in the quasi-linear regime
favors colinear structures (see Figure \ref{fig:tri}), 
which can be identified,
in the early stages of structure formation, with collapsed
structures such as sheets and filaments.
The decrease in $Q_V$ with increasing $u$ is due to a combination of
the scale and shape dependences; larger values of $u$ pick out
both larger and more elongated structures, giving rise to competing
effects. Under our chosen normalization convention, the former
effect wins out. 
Fry (1994, 1996) finds similar results by 
considering
the $k$-space $\mbox{${\cal Q}$}$, defined in equation~(\ref{B}).
Though the CDM models shown all exhibit these common trends,
the precise behaviors predicted for $Q_V$ by models with
different values of $\Gamma$ are quite
distinct, owing to the different locations of the peaks
in their power spectra. 
Overall, Figure \ref{fig:zeta_actual} suggests that 
measurements of the 3PCF on these scales, though not very sensitive
to changes in the value of $n$ (see BKJ), may discriminate well between
high-$\Gamma$ and low-$\Gamma$ models.

\subsection{Effects of Bias and Its Evolution}

\begin{figure}[htbp]
\plotone{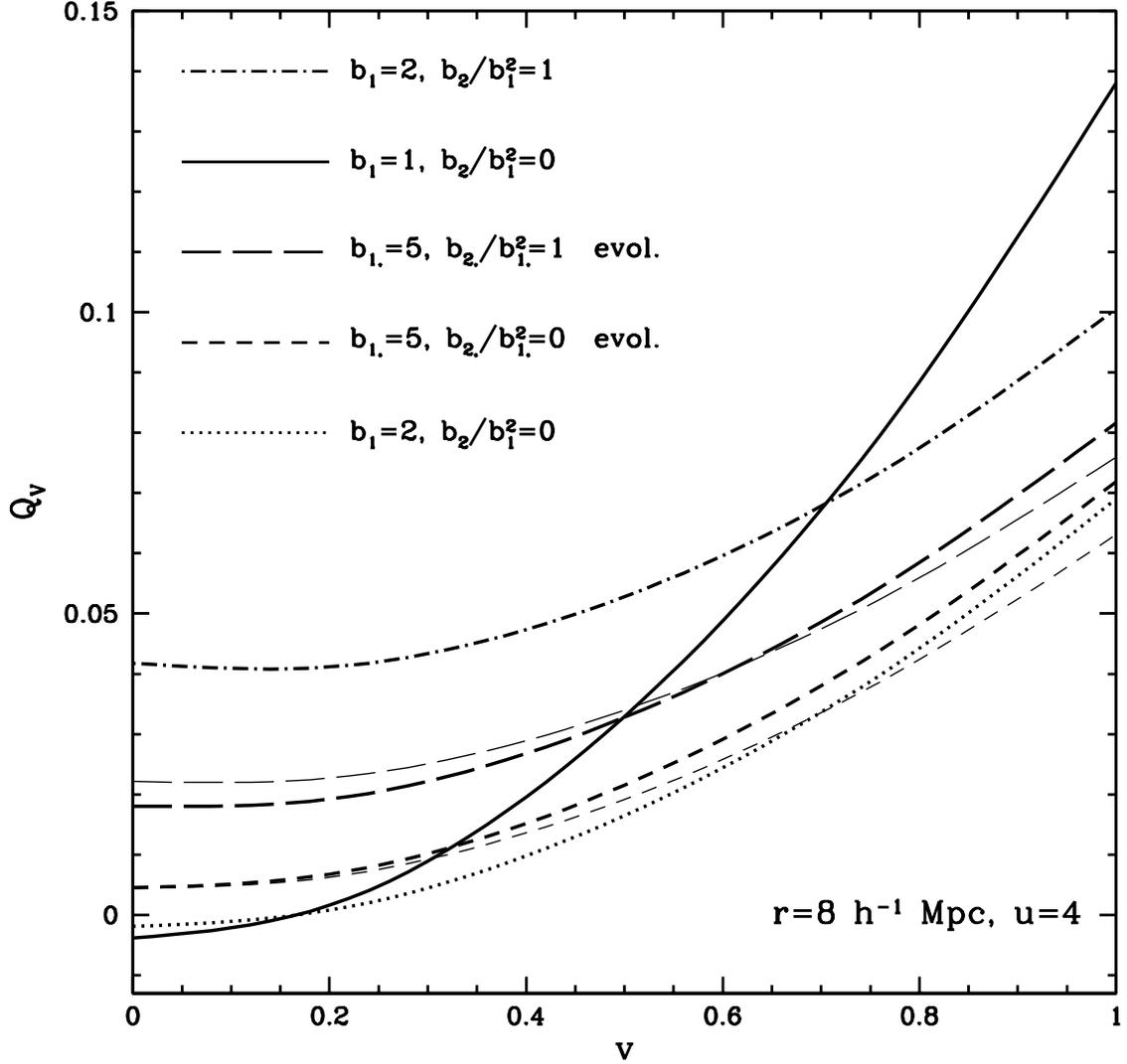}
\caption{Predicted $Q_V(v)$ for $r=8$ $h^{-1}$ Mpc and $u=4$ for 
the five bias 
scenarios shown. For each scenario, the $\Lambda$CDM and OCDM models
are shown in thick and thin lines, respectively, with the 
two models being effectively 
identical in the non-evolving cases, but distinct in the
evolving cases due to their different linear growth factors.}
\label{fig:zetabias}
\end{figure}

Figure \ref{fig:zetabias} shows $Q_V(v)$ for the five bias scenarios from
\S2.2 for the
OCDM and $\Lambda$CDM models (taken to be
identical in the non-evolving bias cases) for $r=8$ $h^{-1}$ 
Mpc and $u=4$. This represents only a small region of the parameter space
we are exploring, but we find similar dependences
of the 3PCF on the bias scheme as were seen for $S_3$,
namely a general flattening and reduction of $Q_V$ with increasing linear 
bias,
and a relative increase with increasing nonlinear bias as compared
with the corresponding linear scheme.  
In the evolving cases, the difference between these two models,
which arises from their different linear growth factors, would be greater
for lower values of $\Omega_0$.

\begin{figure}[htbp]
\plotone{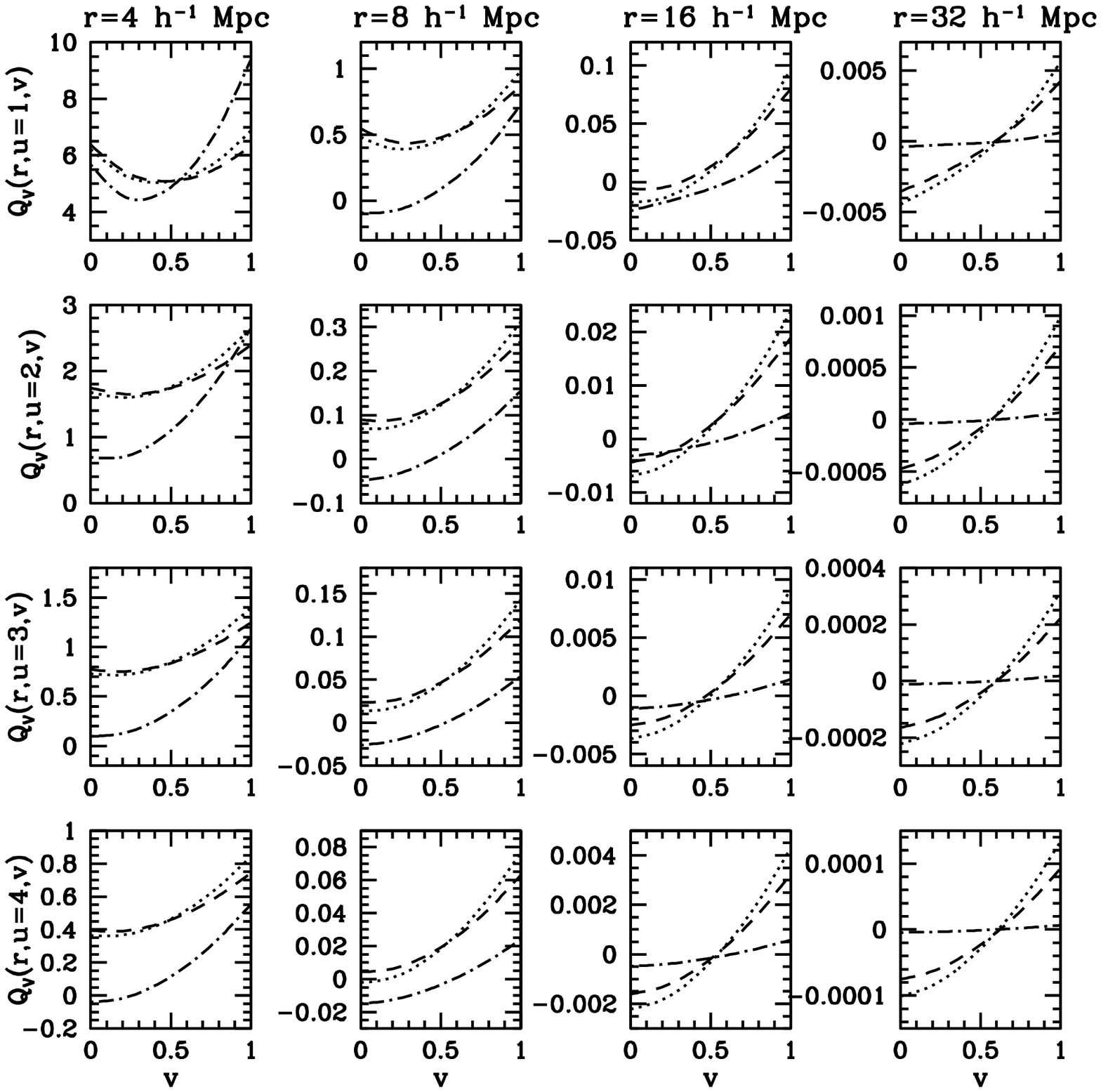}
\caption{Predicted $Q_V(r,u,v)$ for the SCDM, $b_1=2.5$, 
$b_2/b_1^2=0.21$ model
(dot-dashed curves), OCDM, $b_{1_{*}}=5$, $b_{2_{*}}/b_{1_{*}}^2=0$ model
(dashed curves), and
OCDM/$\Lambda$CDM, $b_1=1.8$, $b_2/b_1^2=0$ model (dotted curves).
The constant linear- and nonlinear- bias models (dotted and dot-dashed,
respectively) yielded degenerate $S_3$ predictions,
but can be distinguished using the full geometric dependence of $Q_V$ (in
part because of their different values of $\Gamma$). The
evolving and non-evolving linear models
(dashed and dotted, respectively), however, are not as well distinguished
by measurements of the present-day 3PCF alone.}
\label{fig:zetabiasdeg}
\end{figure}

Whereas the skewness
gave only a single relationship, $S_3(R)$, with which to distinguish
different cosmological and biasing schemes, the full 3PCF allows much stronger
constraints from consideration of the geometric dependence
[i.e., a graph such as Figure \ref{fig:zetabias} for every pair of values
$(r,u)$].  We illustrate this in Figure \ref{fig:zetabiasdeg},
where $Q_V(r,u,v)$ is
plotted for the same three (nearly degenerate) models depicted in Figure
\ref{fig:S3biasdeg}.
The constant linear- and nonlinear-bias schemes (dotted and
dot-dashed curves, respectively), which gave nearly
identical results for $S_3$, yield distinct predictions for $Q_V$,
in part because of their different values of $\Gamma$ as well. It is 
well-known that measurements of the bispectrum and 3PCF
can be used to obtain independent constraints on the constant-valued
linear and nonlinear bias parameters 
(FG93; Fry 1994, 1996; Jing 1997; Matarrese, Verde, \&
Heavens 1997); this can be seen here from the different manners in which
these terms appear in equation~(\ref{zeta2}).
The low-$\Gamma$ models in Figure \ref{fig:zetabiasdeg}, 
with linear, evolving (dashed curves)
and non-evolving (dashed curves) bias, however, remain fairly similar, 
since
an initially large, evolving value of $b_1$ again approximates 
a correspondingly 
smaller constant term. Though the relative differences between
these models are slightly more pronounced at larger scales, it is
unlikely that current and emerging redshift surveys will attain
sufficient precision to resolve them. Thus, while the geometric information
contained in the present-day 3PCF can, unlike $S_3$, 
be used to distinguish between different
cosmological models and separate the contributions of 
constant linear- and nonlinear-bias terms, 
it may not reliably distinguish between constant and
evolving bias.  

\begin{figure}[htbp]
\plotone{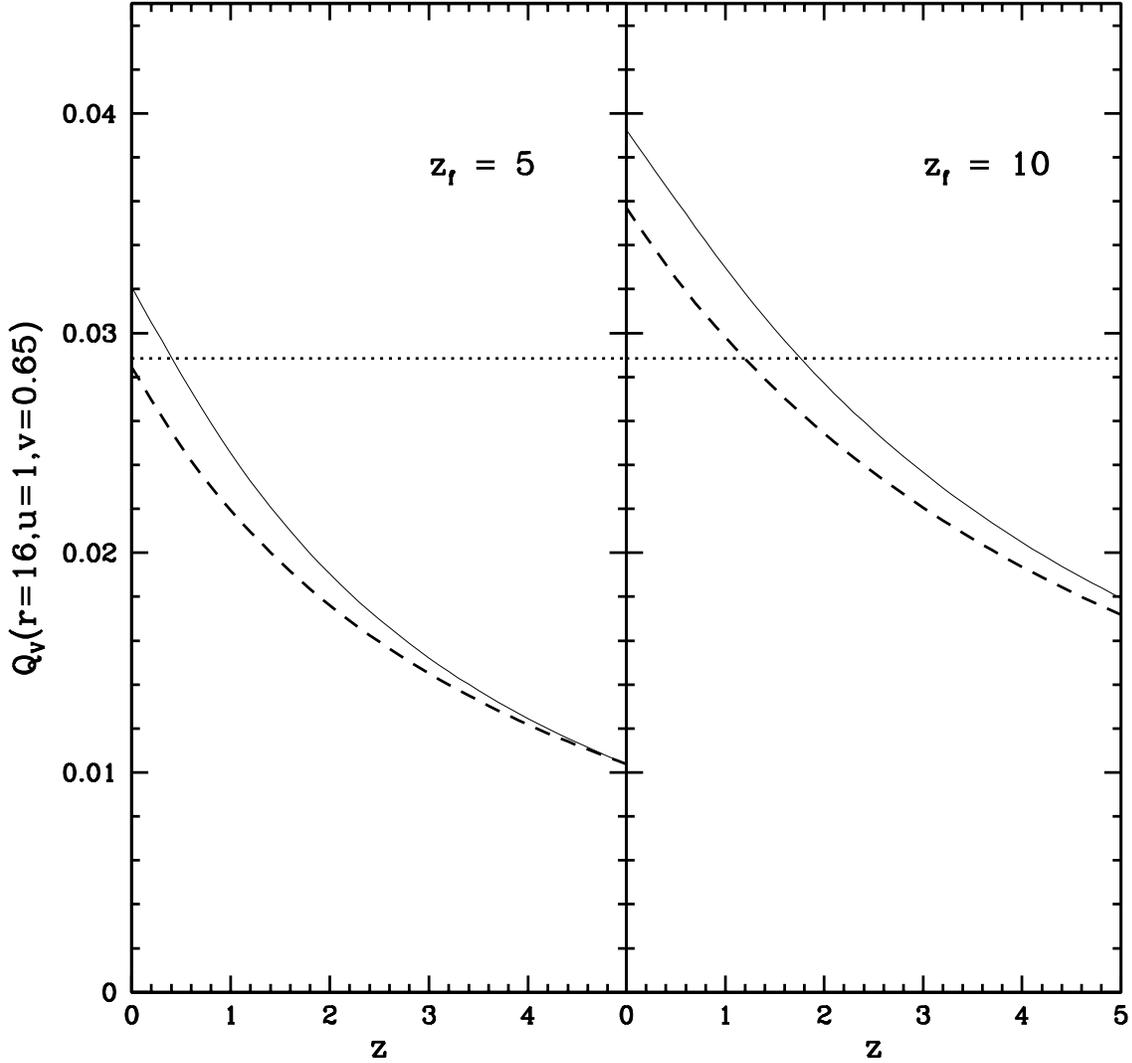}
\caption{The variation of $Q_V$ with redshift
for a fixed triangle configuration ($r=16$ $h^{-1}$ Mpc, $u=1$, $v=0.65$).
The dotted lines show the result for the OCDM/$\Lambda$CDM model
with constant
linear bias, $b_1 = 1.8$, while the dashed 
lines show the evolving
linear bias, $b_{1_{*}}=5$, result for the OCDM. 
The evolving and constant OCDM models are virtually identical
at $z=0$, but can be distinguished by their different redshift-dependences.
An evolving, $\Lambda$CDM model is shown by the thin, solid lines, for
comparison.
The right panel shows the results obtained using a formation redshift
of $z_f=10$, rather than $z_f=5$.}
\label{fig:zetabiasdegofz}
\end{figure}

This ambiguity could, in principle, be removed by considering measurements
of the spatial 3PCF as a function a redshift, as might be obtained
from very deep surveys. Figure \ref{fig:zetabiasdegofz} shows the evolution
of $Q_V$ with redshift, for a particular configuration 
($r=16$ $h^{-1}$ Mpc, $u=1$, $v=0.65$), for the 
OCDM evolving linear bias model
and OCDM/$\Lambda$CDM constant
linear bias model from Figures 
\ref{fig:S3biasdeg} and \ref{fig:zetabiasdeg},
with a $\Lambda$CDM evolving linear
bias model added for comparison.
These evolving and non-evolving cases yielded similar results for
$Q_V$ as well as $S_3$, and for this particular configuration were
practically identical, as can be seen in the appropriate panel
of Figure
\ref{fig:zetabiasdeg}, or from the $z=0$ result in 
Figure \ref{fig:zetabiasdegofz}.
We see, however, that the constant linear bias model can be distinguished
from the evolving models on the basis of their evolution with
redshift. Note that, unlike the constant-bias case where the
negligible
dependence on $\mu$ caused the OCDM and $\Lambda$CDM predictions
to differ only by the thickness of the curves,
the evolving OCDM and $\Lambda$CDM models here yield different results, 
as expected, with the differences becoming larger
for smaller values of $\Omega_0$. Comparing the left and right panels
further illustrates the dependence
on the redshift of galaxy formation, $z_f$, which, in principle,
can also be constrained.

In theory, a survey large and deep enough to resolve this degeneracy
between constant and evolving bias might also enable a model-dependent
determination
of $b_1$ and $b_2$ solely from measurements of $S_3$, as described in
\S2.1. Since distinctions between evolving and non-evolving bias,
such as those drawn in Figure \ref{fig:zetabiasdegofz}, could also
be made by considering the redshift dependence of $S_3$, we can infer
that such a survey might address the various degeneracies between
$b_1$, $b_2$, and their evolution, solely from measurements of the skewness,
without the need to consider the configuration dependence of the full 3PCF.
Unfortunately, it is unlikely that 
upcoming redshift surveys will be both deep and
large enough to resolve the degeneracy of $\zeta$ with respect to
constant vs. evolving bias, and in any case  
would still be susceptible to redshift distortions which mix the
density and velocity fields. However,
BKJ show that the angular 3PCF, which can be well-sampled over
a large range of redshifts (e.g., by radio surveys) and
is free of such distortions, might be better able 
to discriminate between evolving and non-evolving bias scenarios.

Based on these results, it is clear
that the often-stated result that $Q$ is a constant of order
unity (Groth \& Peebles 1977;  Jing \& Zhang 1989;
T\'{o}th, Holl\'{o}si, \& Szalay 1989; Gott, Gao, \& Park 1991;
Jing, Mo, \& B\"{o}rner 1991; Baumgart \& Fry 1991; Bouchet \etal 1993) 
is only a crude approximation resulting from
a coarse averaging over allowed configurations and systematic errors
arising from the normalization convention. Valuable
information is lost if the geometric dependence of $Q$, or
more appropriately $Q_V$, is ignored. 
We also stress that 
Figures \ref{fig:zeta_actual}--\ref{fig:zetabiasdegofz},
are only illustrations;
the statistical leverage gained by considering the full 
configuration-space variation
of the 3PCF as a function of redshift
should in principle discriminate between the signatures
of gravitational evolution versus bias, independently measure the
effective present-day values of $b_1$ and $b_2$, the
shape of the power spectrum, and possibly measure the evolution in these
quantities and constrain the epoch of galaxy formation.

\section{CONCLUSION}

We have derived leading-order results for the normalized spatial
skewness and 3PCF, assuming Gaussian ICs, 
for an arbitrary tracer-mass distribution in
flat and open universes, taking into account such features
as the scale dependence and linear evolution of the CDM power spectrum 
and the presence of a possibly evolving (decaying),
linear or nonlinear bias, as defined for unsmoothed density fields. 
The predicted normalized amplitudes, $S_3$ and $Q$,
for an unbiased tracer mass are in agreement with previous work 
(Bouchet \etal 1992; Juszkiewicz, Bouchet, \& Colombi 1993;  
FG93; Bernardeau 1994a; Fry 1994; JB97).
We extend the FG93 result for $S_3$ in the case of a 
nonlinear bias, defined for unsmoothed fields, to include a
scale-dependent, leading-order correction which becomes appreciable for
positive effective spectral indices, corresponding to scales
$R \ga 100$ $h^{-1}$ Mpc for CDM models. This correction term
implies that the value of the nonlinear bias parameter, as defined
for smoothed density fields, could generally
depend on the adopted smoothing scale. 
In the unlikely event
that $S_3$ could be measured over 
scales $R \ga 100$ $h^{-1}$ Mpc, the presence or absence of this behavior
could shed light on the smoothed vs. unsmoothed biasing
picture, and perhaps allow more accurate determinations
of the linear- and nonlinear-bias parameters on the basis of skewness
measurements alone.  

We show that the conventional definition of
$Q$ gives rise to rapid, non-monotonic variation and divergences
which do not arise from
the behavior of the 3PCF.
We speculate that the large discrepancy between this behavior and that
predicted by some $N$-body simulations is due 
to a practical bias, arising from the poor normalization
convention, in the reconstruction of $Q$ from data, 
rather than from the absence of nonlinear corrections in the QL PT
prediction. We propose that this question may be addressed by
considering $N$-body results at earlier cosmic epochs.
We consider instead $Q_V$, which is everywhere well-behaved.

We find the bias model to be a crucial factor influencing the
clustering predictions, in agreement with Matarrese \etal (1997).
In particular, the scale dependence of $S_3$ and the
configuration dependence of $Q_V$ bear characteristic
imprints of bias, such as a relative flattening and decrease with increasing
$b_1$ and a relative increase with increasing $b_2$. 
Unlike $S_3$, the predictions for $Q_V$ are seen to depend significantly
on the adopted CDM model, through the dependence on $\Gamma$, 
and for all models investigated, the 
dependence of the 3PCF on triangle geometry 
implies stronger clustering of smaller, elongated structures.
The statistic $S_3$, while preserving information
about the overall scale dependence, is seen to provide only one
test on the combination of the cosmological parameters,
$b_1(t)$, and $b_2(t)$, yielding
degeneracies between these quantities, particularly for smoothing
scales $R < 100$ $h^{-1}$ Mpc.  These are partially alleviated
by considering the dependence of the full 3PCF
on the triangle geometry. In particular, the variation
of $Q_V$ can be used to distinguish between the effects
of gravitational evolution and bias, place
constraints on the value of $\Gamma$, 
and measure
the effective constant bias terms $b_1$ and $b_2$.   
Evolving-bias models, however, are found to yield similar predictions
for both $Q_V$ and $S_3$ as models with an appropriately smaller,
constant bias, and thus cannot
be reliably distinguished using spatial statistics at $z=0$.
Measurements of the 3PCF as a function of redshift could, in
principle, directly measure bias evolution, $\Omega_0$,
and the redshift of galaxy formation, but these might
be more readily measurable using angular statistics (BKJ).
A comparison of these predictions for $S_3$ and $Q_V$
with several data sets 
characterizing different tracer populations would allow
multiple, complementary constraints on the above parameters.
The often-quoted ``empirical'' result, that the normalized
3PCF is a constant of order unity,
is in part a result of coarse-averaging over all triangular configurations, 
which destroys
much of this valuable information.  
  
In practice, other factors arise when comparing these predictions
with observations. For example, redshift distortions, which mix information
about the density and velocity distributions, must be accounted for
when measuring spatial clustering
(Fry \& Gazta\~{n}aga 1994; Jing \& B\"orner 1998; Verde \etal
1998; Heavens, Matarrese, \& Verde 1998; Scoccimarro, Couchman,
\& Frieman 1998). Finite-volume
and boundary effects, as well as estimation biases, tend to
reduce the observed clustering amplitude (Baugh, 
Gazta\~{n}aga, \& Efstathiou 1995; Szapudi \& Colombi 1996;
Colombi, Szapudi, \& Szalay 1998; Gazta\~{n}aga \& Bernardeau 1998; Hui \&
Gazta\~{n}aga 1998), and Poisson noise (Gazta\~{n}aga 1994) and sampling
variance must also be taken into account. 
In addition, the calculations above can
be extended in various ways. Our assumed model for the bias is 
deterministic, but other stochastic models have been proposed
which may yield different results, particularly on smaller scales 
(FG93; Catelan \etal 1998; Catelan,
Matarrese, \& Porciani 1998; Taruya, Koyama \& Soda 1998). 
One might also investigate the effects of including higher-order
nonlinear terms in the PT expansion, such as one-loop results
(Jain \& Bertschinger 1994; Scoccimarro \& Frieman 1996a,b;
Scoccimarro \etal 1998);  these are expected to be negligible
on QL scales, but become increasingly important when comparing
predictions with
observations on scales where $\xi \ga 1$. For example,
$N$-body simulations indicate that $Q$ becomes relatively shape-independent
in the nonlinear regime, suggesting that a linear bias is not the only
mechanism capable of producing a flattening (Scoccimarro \etal 1998), and
Gazta\~{n}aga \& Bernardeau (1998) also find that nonlinear
effects tend to increase $S_3$ while erasing the shape dependence of $Q$.
Calculations analogous to those presented here can also
be performed for higher-order moments
and $n$-point CFs (e.g., kurtosis, four-point correlation function, etc.)
which, in principle, should provide additional constraints.
Finally, the entire discussion can be extended to include
the predictions of structure-formation
models with non-Gaussian ICs, such as topological-defect or
isocurvature models (e.g., Jaffe 1994).

\bigskip
We are grateful to Roman Scoccimarro for numerous discussions and
useful suggestions. We also wish to thank
Licia Verde, as well as the anonymous referee,
for providing helpful comments. This work was supported 
by a D.O.E. 
Outstanding Junior Investigator Award, DEFG02-92-ER 40699, NASA
NAG5-3091, NSF AST94-19906, and the Alfred P. Sloan Foundation.

\end{document}